\documentclass[aps,prd,showpacs,preprintnumbers,nofootinbib,floatfix,onecolumn]{revtex4}

\usepackage{bm}
\usepackage{latexsym}
\usepackage{dcolumn}
\usepackage{amsfonts,amssymb}
\usepackage{graphicx,epsfig}
\usepackage{psfrag}
\usepackage{subfigure}

\def\beq{\begin{equation}}
\def\eeq{\end{equation}}
\def\br{\begin{eqnarray}}
\def\er{\end{eqnarray}}
\def\pa{{\partial}}
\def\l{\left}
\def\r{\right} 
\def\nn{\nonumber}   
\def\eq#1{{Eq.~(\ref{#1})}}

\begin{document}
  

\title{Particle creation, classicality and related issues in\\ quantum field theory: II. Examples from field theory}
\author{Gaurang Mahajan}
\email[]{gaurang@iucaa.ernet.in}
\affiliation{IUCAA, Post Bag 4, Ganeshkhind, Pune - 411 007, India\\}
\author{T. Padmanabhan}
\email[]{nabhan@iucaa.ernet.in}
\affiliation{IUCAA, Post Bag 4, Ganeshkhind, Pune - 411 007, India\\}

\date{\today}


\begin{abstract} 
We adopt the general formalism, which was developed in Paper I (arXiv:0708.1233) to analyze the evolution of a quantized time-dependent oscillator, to address several questions in the context of quantum field theory in time dependent external backgrounds. In particular, we study the question of emergence of classicality in terms of the phase space evolution and its relation to particle production, and clarify some conceptual issues. We consider a quantized scalar field evolving in a constant electric field and in FRW spacetimes which illustrate the two extreme cases of late time adiabatic and highly non-adiabatic evolution. Using the time-dependent generalizations of various quantities like particle number density, effective Lagrangian etc. introduced in Paper I, we contrast the evolution in these two limits bringing out key differences between the Schwinger effect and evolution in the de Sitter background. Further, our examples suggest that the notion of classicality is multifaceted and any one single criterion may not have universal applicability. For example, the peaking of the phase space Wigner distribution on the classical trajectory \emph{alone} does not imply transition to classical behavior. An analysis of the behavior of the \emph{classicality parameter}, which was introduced in Paper I, leads to the conclusion that strong particle production is necessary for the quantum state to become highly correlated in phase space at late times.        
\end{abstract}

\maketitle


\section{Introduction and Motivation}

The subject of quantum field theory in external classical backgrounds has received significant attention in the literature~\cite{txts1,txts2,schwinger,itz,efield,parker}. Though the mathematical treatment is relatively straightforward, several conceptual issues remain to be settled. This paper is a sequel to a previous one in which we attempted to clarify some of these open issues in the context of a single oscillator with general time dependent parameters. Here, we shall shift our attention to the broader context of quantum fields evolving in certain time dependent external backgrounds, and in particular try to  understand the relation of particle production with the approach to classicality in terms of phase space correlations. 

We  begin by recalling the basic issues that were raised (and elaborated on) in Paper I~\cite{gm}, that one encounters in any semiclassical analysis:
(1) In a general time dependent background, there is no well-defined procedure for defining the notion of particles. If adiabatic regions cannot be defined asymptotically, the usual definition of particles in terms of \emph{in} and \emph{out} states is rendered invalid. 
(2) A time-dependent particle concept is also required for a proper treatment of the question of back reaction of the quantum system on the classical background, which is not an asymptotic notion. many such definitions for `particles' can be given; our aim is not to come up with a \textit{unique} definition but a \textit{useful} definition and explore its consequences.
(3) The standard treatment for computing backreaction in the semiclassical approximation considers only the real part of the effective Lagrangian in determining the modified background equations of motion. The non-zero imaginary part (which directly quantifies the asymptotic particle content) is usually ignored in such a calculation, because the resulting equations (and solutions) will be complex (see for e.g. ref.~\cite{brown}). This leads one to ask whether information about particle production gets encoded in some way in the real part. 
(4) The concept of particles is intimately tied with the notion of classicality of the state. This issue has particular relevance in studying the generation of primordial perturbations in inflationary cosmology~\cite{infl1,infl2,paddy03,cosmo,ps}: in the standard approach, these density perturbations are assumed to be generated as vacuum fluctuations of a quantized (inflaton) field, but their late time evolution at super-Hubble scales is treated using classical notions, with the only reference to their quantum origin being in the choice of initial conditions. This naturally brings up the question of precisely how something that originates as a quantum mechanical entity can develop classical features over time.

In Paper I, we confined ourselves to the case of a quantized time dependent oscillator and presented a general formalism to quantify the time dependent physical content of an evolving Gaussian quantum state in the Schrodinger picture. In particular, we chose to adopt an \emph{instantaneous} concept of a particle, defining it in terms of instantaneous eigenstates specified at each moment in close analogy with a simple time-independent oscillator. (We emphasize that we view this as a useful definition rather than a unique definition.) As illustrated by its subsequent application to several toy examples, this approach provides a physically reasonable picture of the time evolution, with reference to particle content and phase space correlations. The main conclusions reached by our analysis of an oscillator with a general frequency $\omega(t)$ were: (1) Peaking of the Wigner function on the classical phase space trajectory (usually associated with classicality) is a rather general feature that can happen even when the oscillator evolves adiabatically, i.e. it is quite independent of the particle production. (2) To make the idea of interpreting classicality less ambiguous, we  suggested considering an additional correlation measure (calling it the classicality parameter ${\cal S}$); this quantity was found to track the particle creation more closely, and it does not always follow the behavior of the Wigner distribution. Based on an approximate asymptotic analysis of a general $\omega(t)$, we identified two distinct late time limits: whenever the oscillator evolves adiabatically ($|\epsilon(t)| \ll 1$) at late times, particle creation is suppressed and the classicality parameter maintains an oscillatory profile; in contrast, when adiabaticity is violated (with $|\epsilon(t)| \to \infty$), strong particle production occurs and this is accompanied by a sharp growth in ${\cal S}$, whose magnitude saturates at the maximum value of unity. 

Here, we shall exploit the above  ideas in the semiclassical context of quantum fields in time dependent external classical backgrounds, laying particular emphasis on making precise the connection of particle production with the emergence of classicality in the phase space picture. We will take up two very conventional examples, that of a massless minimally coupled scalar field in an FRW spacetime and a complex scalar field in a constant electric field background, which will prove to be quite illuminating for our purposes. (These are, of course, very well-studied examples which allows us one to make comparison with previous results.) The form of the action functional  in these contexts allows, through a Fourier decomposition, to decompose the field as a set of uncoupled harmonic oscillators, with the time dependence of the background variables getting incorporated in the oscillator parameters. This reduces the field theory problem to one studied in Paper I and our formalism becomes directly applicable. (There is, of course, the standard issue of divergences which arise when one adds up infinite number of oscillators. While this is important in a different conceptual context, it is irrelevant for our purposes when we will be dealing with a quadratic action in a given background. Alternatively, one can imagine that a cut-off is imposed in the Fourier space at high momenta and attention is confined to modes well below the cut-off.)

This paper is structured as follows. Section~\ref{sec:formalism} provides a succinct outline of the key ideas of the formalism that was developed in detail in Paper I~\cite{gm} to analyze a general time-dependent oscillator, placing it now in the broader context of the field picture. Section~\ref{sec:examples} is devoted to a detailed analysis, in Fourier space, of the evolution of a minimally coupled real scalar field in expanding FRW spacetimes and of a complex scalar field in a constant electric field background, with particular attention given to particle production and the phase space evolution of the Wigner function of the quantized field modes. Our results essentially corroborate and strengthen the conclusions that were drawn on the basis of the toy models we analyzed in Paper I, in relation to the dependence of phase space correlations on the particle production (i.e. adiabatic vs. non-adiabatic evolution). This is followed by a discussion and conclusion in section~\ref{sec:diss}.           
In what follows, we shall set $\hbar=c=1$.

\section{Formalism}  \label{sec:formalism}

As mentioned in the introduction, we  consider here scenarios involving a quantized field, say $\Phi({\bf x},t)$, coupled to an external classical background where the action for the field is expressible, by means of a Fourier transform, in the following form:
\beq
{\cal A}[\Phi] = \frac{1}{2}\int d^{3}{\bf k}\int dt\; m_{\bf k}(t)\,
\l(\dot{q}_{\bf k}^{2} - \omega^2_{\bf k}(t)\, q_{\bf k}^2\r).    \label{gen_actn}
\eeq
Here, $q_{\bf k}$ is a real variable associated with the Fourier mode ${\bf k}$ and stands for the coordinate of the ${\bf k}$th oscillator. $t$ is the time coordinate and the dot denotes differentiation w.r.t. $t$. The oscillator variables $m_{\bf k}$ and $\omega_{\bf k}$ of each Fourier mode incorporate the time-dependence of the external background. (To be more precise, if the background is characterized by time dependent parameters collectively denoted by $C(t)$, the mass and frequency, being assumed to be functions of $C$, are implicitly time-dependent.) The form of \eq{gen_actn} allows for a mode-by-mode analysis of the quantized field to be carried out, once the functions $C(t)$ are specified.

Although we will not go into the details here (for which the reader is referred to Paper I~\cite{gm}), we provide a quick summary of the Schrodinger picture formalism to study the time evolution of the time dependent oscillator, to set up the notation.
We will consider form invariant Gaussian states with vanishing mean~\cite{pad,gauss,tpgswfn} to describe each oscillator, which have the general form
\beq
\psi (q_{\bf k},t) = N \exp[- R_{\bf k}(t) q_{\bf k}^{2}] \equiv N\exp \l[ - \frac{m_{\bf k}\omega_{\bf k}}{2} \l(\frac{1-z_{\bf k}}{1+z_{\bf k}} \r) q_{\bf k}^{2} \r]    \label{gwfn}
\eeq
where the wave function has been  written equivalently in terms of the either the function $R_{\bf k}(t)$ or $z_{\bf k}(t)$. Using the time dependent Schrodinger equation, these functions can be shown to satisfy first order (but nonlinear) differential equations:
\beq
\dot{R}_{\bf k} = \frac{2 R_{\bf k}^2}{m_{\bf k}} -\frac{1}{2}m_{\bf k}\omega_{\bf k}   \label{eqR}
\eeq
and
\beq
\dot z_{\bf k} + 2 i \omega_{\bf k} z_{\bf k} + \frac{1}{2} \l(\frac{\dot \omega_{\bf k}}{\omega_{\bf k}} + \frac{\dot m_{\bf k}}{m_{\bf k}} \r) (z_{\bf k}^{2} - 1) = 0,  \label{eqz}
\eeq
and we shall denote the quantity $((\dot{m_{\bf k}}/m_{\bf k}) + (\dot{\omega_{\bf k}}/\omega_{\bf k}))/\omega_{\bf k}$, called the adiabaticity parameter, as $\epsilon_{\bf k}(t)$. The problem of determining the quantum evolution thus boils down to directly solving for $R_{\bf k}$ or $z_{\bf k}$ (with an appropriate choice of initial conditions). Alternatively, since \eq{eqR} and \eq{eqz} are of the generalized Riccati type, they can be transformed into second order linear equations; for example, setting $R_{\bf k}=-\l(i\, m_{\bf k}/2\r)(\dot \mu_{\bf k}/ \mu_{\bf k})$ where $\mu_{\bf k}(t)$ is a new function in \eq{eqR} implies that $\mu_{\bf k}$ satisfies the following differential equation:
\beq
\ddot \mu_{\bf k} + \frac{\dot m_{\bf k}}{m_{\bf k}}\, \dot \mu_{\bf k}
+\omega_{\bf k} ^2 \mu_{\bf k}=0
\label{mueq}
\eeq
which is same as the classical equation of motion satisfied by the oscillator variable $q_{\bf k}$, and the Wronskian condition $\dot{\mu}_{\bf k}^{*}\mu_{\bf k} - \dot{\mu}_{\bf k}\mu_{\bf k}^{*} = -iW({\bf k})/m_{\bf k}$, where $W({\bf k})$ is independent of time, also holds. In some cases, it turns out to be more convenient to solve \eq{mueq} rather than start from \eq{eqz}. But this obviously makes no real difference as $z_{\bf k}$ is trivially related to $\mu_{\bf k}$:
\beq
z_{\bf k} = \l(\frac{\omega_{\bf k}\mu_{\bf k} + i \dot{\mu}_{\bf k}}{\omega_{\bf k}\mu_{\bf k} - i \dot{\mu}_{\bf k}}\r).
\eeq
Although $\mu_{\bf k}(t)$ (or $z_{\bf k}$) completely determines the state of the system, it is useful, particularly from the point of view of understanding the physical content of the evolving quantum state, to define a set of variables with physically reasonable interpretation, that can be built out of the wave function. For this purpose, following our approach of Paper I~\cite{gm}, we consider the following quantities:

(1) The mean particle number $\langle n_{\bf k} \rangle$, that specifies for us the instantaneous particle content of the quantum state and is defined using the concept of instantaneous eigenstates. The mean particle number as well as the mean energy in a given mode (defined as the expectation value of the Hamiltonian) at any moment $t$ are specifiable in terms of $z_{\bf k}$:
\beq
\langle n_{\bf k} \rangle = \frac{|z_{\bf k}|^2}{1-|z_{\bf k}|^2} \quad;\quad {\cal E}_{\bf k} = \l( \langle n_{\bf k} \rangle + \frac{1}{2} \r) \omega_{\bf k}(t). 
\eeq   

(2) The time-dependent generalization of the effective Lagrangian $L_{eff}({\bf k},t)$~\cite{schwinger,itz,infl2}, that specifies the transition amplitude between instantaneous vacuum states defined at two different moments, which again is related in a simple manner to $z_{\bf k}$:
\beq
L_{eff}({\bf k},t) = \frac{i}{4} \l( \frac{\dot \omega_{\bf k}}{\omega_{\bf k}} + \frac{\dot m_{\bf k}}{m_{\bf k}} \r) z_{\bf k}.  \label{L_eff}
\eeq
This is to be understood as the contribution of the ${\bf k}$th mode to the effective Lagrangian for the field. (This expression is somewhat formal since the functional dependence of $L_{eff}({\bf k},t)$ on the brackground variables is not explicit in this; in spite of this, we will see that it is useful.)

(3) The Wigner distribution function defined in the $q_{\bf k}$-$p_{\bf k}$ phase space of the oscillator, that provides a reasonable means of quantifying the approach to classicality of the quantum state in terms of its phase space evolution~\cite{infl2,wig,pad,gauss}. The Wigner function for the Gaussian state in \eq{gwfn} can be written as
\beq
 {\cal W}\l(q_{\bf k}, p_{\bf k}, t\r) = \frac{1}{\pi} \exp\l[-\frac{q_{\bf k}^2}{\sigma_{\bf k}^2(t)}-\sigma_{\bf k} ^2(t)\, \l(p_{\bf k}-{\cal J}_{\bf k}(t)\,  q_{\bf k}\r)^2 \r]
 \label{wigfun}
\eeq 
where
\beq
\sigma_{\bf k} ^2 = \frac{|1+z_{\bf k}|^2}{m_{\bf k} \omega_{\bf k}(1-|z_{\bf k}|^2)} = \frac{2|\mu_{\bf k}|^2}{W}
\eeq
and
\beq
{\cal J}_{\bf k} = \frac{2 m_{\bf k} \omega_{\bf k} Im(z_{\bf k})}{|1+z_{\bf k}|^2} = \l(\frac{m_{\bf k}}{2}\r) \frac{1}{|\mu_{\bf k}|^2} \frac{d |\mu_{\bf k}|^2 }{d t}.
\eeq
In addition, we define an additional measure of phase space correlations, calling it the classicality parameter ${\cal S}$, which is given by
\beq
{\cal S}_{\bf k} = \frac{{\cal J}_{\bf k} \sigma_{\bf k}^2}{\sqrt{1+ \l({\cal J}_{\bf k} \sigma_{\bf k}^2 \r)^2}}.
\label{classicality}
\eeq
The classicality parameter vanishes for a vacuum state, and its value is confined to the interval $[-1,1]$. (The quantity ${\cal J}_{\bf k} \sigma_{\bf k}^2$ has the simple interpretation of being just twice the phase space average of $q_{\bf k} p_{\bf k}$.) Our analysis in Paper I suggested contrasting behavior of the Wigner function and the classicality parameter in different circumstances; the classicality parameter (but not the Wigner function) was shown to closely mirror the variation of $\langle n \rangle$. 

Since one is dealing with a field here, one can go a step further here and choose to look at the evolution in coordinate space which we could not do in the case of single oscillator studied in Paper I. Consider, in particular, the coordinate space representation of the function $z_{\bf k}$ by taking a Fourier transform and defining a function $Z(x^i)$. Working in the Heisenberg picture, it can be shown that $z_{\bf k}$ is related to the transition amplitude between: (i) a vacuum state defined at some initial instant $t_i$ and (ii) a two particle state (with momenta ${\bf k},-{\bf k}$) defined using the instantaneous the vacuum state at time $t$:
\beq
z_{\bf k}(t) = \frac{\langle 0, t |\hat{A}_{-{\bf k}}(t)\hat{A}_{{\bf k}}(t)| 0, t_i \rangle}{\langle 0, t | 0, t_i \rangle} e^{-2 i \rho_{\bf k}(t)}
\eeq
where $\dot{\rho_{\bf k}}(t)=\omega_{\bf k}(t)$ and $\hat{A}_{{\bf k}}(t)$ is the annihilation operator defined at the instant $t$ (see Paper I~\cite{gm} for a derivation of this relation). Multiplying both sides by $\exp(i{\bf k \cdot r})$ and integrating over ${\bf k}$, we get
the real space function: 
\beq
Z({\bf r},t)  \equiv \int z_{\bf k} e^{i {\bf k \cdot r}} \frac{d^3 {\bf k}}{\l( 2 \pi \r)^3}. 
\eeq
This quantity has an interesting physical interpretation. If we define the 
 positive frequency part of the field by 
 \beq
\hat{F}({\bf x},t) = \int \hat{A}_{{\bf k}}(t) e^{i {\bf k \cdot x} - i\rho_{\bf k}} \frac{d^3 {\bf k}}{\l(2 \pi \r)^{3/2}}, 
\eeq
then the state $ |{\bf x}, {\bf y}; t \rangle=\hat{F}^\dagger({\bf x},t) \hat{F}^\dagger({\bf y},t)|0,t\rangle$ can be thought of as a state with two particles at ${\bf x}$ and ${\bf y}$.
The quantity $Z({\bf r},t)$ can be expressed as
\beq
Z({\bf x}-{\bf y},t) = \frac{\langle 0, t |\hat{F}({\bf x},t) \hat{F}({\bf y},t)| 0, t_i \rangle}{\langle 0, t | 0, t_i \rangle} \equiv \frac{\langle {\bf x}, {\bf y}; t |0, t_i \rangle}{\langle 0, t | 0, t_i \rangle}.
\eeq
 This quantity  can, therefore, be interpreted as the amplitude for the initial vacuum state to appear as a state with two particles at points  ${\bf x}$ and  ${\bf y}$ at the moment $t$.  Obviously it contains some information relating to the quantum versus classical nature of the field. One would have expected the particles to be uncorrelated --- so that $Z({\bf x}-{\bf y},t)\to \delta ({\bf x}-{\bf y})$ --- in the classical context, while the two particles created at two different points to retain a long range correlation in the quantum regime. It is therefore worthwhile to analyze the evolution of this quantity in different contexts.

Since all the quantities defined above are related to the real and imaginary parts of $z_{\bf k}$ in a simple manner, it is operationally  convenient to determine $z_{\bf k}$ and then determine the various physical quantities.  Our analysis in Paper I thus focused on solving for the time-evolution of $z$ in different scenarios involving various functional choices for $\omega(t)$. In particular, we considered in detail toy models to illustrate the two kinds of late time evolution possible: for example, a frequency function $\omega_{1}(t) = \lambda \sqrt{1+\lambda^{2} t^{2}}$,  goes over into an adiabatic phase at late times ($\lambda t \gg 1$) with the adiabaticity parameter $\epsilon(t)$ going to zero, while another function  $\omega_{3}(t) = \lambda / (1 + \lambda^{2} t^{2})$, is characterized by a strongly non-adiabatic late time regime (with the magnitude of $\epsilon(t)$ diverging as $t \to \infty$). We found that the late time behavior (starting from an instantaneous vacuum state at a sufficiently early moment), while differing appreciably in general, shows similarities in certain respects. Although the instantaneous mean particle number follows the expected variation, saturating in the adiabatic example but diverging in the non-adiabatic case, as one would have expected, the Wigner distribution ends up peaking on the corresponding classical phase space trajectory in \emph{both} situations, being clearly independent of the nature of the evolution, and is therefore of little use in making a statement about the approach to classicality. This prompted us to also look at the classicality parameter as an additional measure to characterize the phase space correlations; this quantity was found to closely track the particle production, and like the particle number, showed a clear difference in the two limits, remaining bounded (with oscillations) in the adiabatic case but saturating at unity (corresponding to a diverging $q$-$p$ phase space average) in the other extreme limit. Taken together, these features point to potential subtleties involved in interpreting classicality in terms of the phase space picture.

With a view to exploring these ideas further, we shall now proceed to take up some standard examples appearing frequently in field theory in the semiclassical context, which are expected to further illustrate the two kinds of late time evolution mentioned in the previous paragraph. First, we will consider the often studied Schwinger effect in a constant electric background, using a complex scalar field. This scenario can be brought into the form of time-dependent oscillator by a particular choice of  gauge. When expressed in the time-dependent gauge, this scenario becomes mathematically identical to the toy model corresponding to the frequency function $\omega_1(t)=\lambda \sqrt{1+\lambda^{2} t^{2}}$ we analyzed in Paper I, which was characterized by adiabatic evolution at very early as well as very late times. On the other hand, the second set of examples we consider --- evolution of a real scalar field in inflationary as well as decelerating FRW backgrounds ---  will differ in certain respects from the electric field case, and these will be pointed out in the course of our analysis. It may be mentioned that the examples referred to above are  very standard, and in which particle creation has been extensively analyzed in the literature~\cite{txts1,infl2,pad,parker,class}. But here we will be re-examining them in the broader context using the definitions introduced in Paper I which seems to lead to a more unified and clearer picture. 

\section{Analysis of examples}  \label{sec:examples}

\subsection{Quantum evolution in a classical electric field background}   \label{sec:ef}

It has been known for quite a while that a classical electromagnetic field can have non-trivial quantum effects, like pair creation out of the vacuum (the so-called Schwinger effect)~\cite{schwinger,itz,infl2,efield}. A simple model that suffices to illustrate this situation involves a complex scalar field $\phi$ evolving in a spatially uniform electric background. The electric field ${\bf E}(t)$ can be described in a time-dependent gauge corresponding to $A_z \equiv A_z(t)$, with the other components being zero, and the action for a scalar field can then be written as~\cite{schwinger,itz,efield}:
\beq
{\cal A}[\phi] = \frac{1}{2}\int {d^{4}x}~(\partial_{\mu} + iqA_{\mu})\phi~(\partial^{\mu} - iqA^{\mu})\phi^{*} \nonumber\\
\eeq
\beq
= \frac{1}{2} \int {d^3{\bf{k}}} \int dt \left(\vert \dot q _{\bf{k}}\vert ^2 - (k_{n}^2 + m^2 + (k_{z} - q A(t))^2 )\vert q_{\bf{k}}\vert ^2\right)    \label{ef_actn}
\eeq
with $k^2=k^2_{n}+k^2_{z}$. Since the scalar field is complex, each mode $\bf{k}$ is associated with two degrees of freedom (corresponding to the real and imaginary parts of $q_{\bf{k}}$); hence the number of oscillators is twice that in the case of a real field.

Comparing \eq{ef_actn} with the general action in \eq{gen_actn}, it follows that the field can be regarded as a collection of uncoupled oscillators, each with unit mass and a time-dependent frequency $\omega_{\bf k}(t)=\sqrt{k_{n}^2 + m^2 + (k_{z} - q A(t))^2}$. The classical equation for the variable $\mu_{\bf k}$,~\eq{mueq}, takes the form
\beq
\ddot \mu_{\bf k} + \omega_{\bf k}^{2}(t)\mu_{\bf k} = 0.   \label{mu_ef}
\eeq

 We shall choose the case of a constant electric field $E>0$ which can be described by the vector potential $A(t)=-Et~(-\infty < t < \infty)$. This example has been widely studied as a particularly simple demonstration of particle creation effects in semiclassical theory, but here we are also interested in considering the phase space evolution of the quantized field.  

By the transformation $\tau = k_z/\sqrt{qE} + \sqrt{qE} t$, \eq{mu_ef} can be converted to a simpler form:
\beq
\frac{d^{2}\mu_{\bf k}}{d\tau ^{2}} + (\tau^{2} + \lambda_{\bf{k}}) \mu_{\bf k} = 0,
\label{eom}
\eeq
where $\lambda_{\bf{k}} = (k_{n}^2+m^2)/qE$. This corresponds to an oscillator with the time-dependent frequency $\omega_{\bf k}(\tau)=\sqrt{\tau^{2} + \lambda_{\bf{k}}}$. For this frequency function, the adiabaticity parameter given by $\epsilon_{\bf k}(\tau)= \tau/(\lambda_{\bf{k}} + \tau^{2})^{3/2}$ vanishes in the asymptotic limits ($\tau \to \pm \infty$), so \emph{in} and \emph{out} vacuum states can be defined in the adiabatic regions. If one starts out in the \emph{in} vacuum state (defined by $\langle n_{\bf{k}} \rangle \rightarrow 0$ as $\tau \to -\infty$), the non-equivalence of the two asymptotically defined vacua implies that the state will appear populated with quanta measured with respect to the \emph{out} vacuum at late times. 

As a first step, we would like to understand how the instantaneous particle number, introduced in Paper I, evolves with time. This quantity is, of course,  expected to coincide with the definition based on in-out states in the adiabatic regions. A computation of the particle content requires one to solve for $z_{\bf k}$ or $\mu_{\bf k}$ and, for the constant electric field case,  \eq{eom} is identical in form to the Schrodinger equation for an inverted harmonic oscillator, which is exactly solvable. We can thus obtain the analytic solution, and this will allow us to determine all the physical properties, including the exact asymptotic particle content at late times.

The linearly independent solutions of \eq{eom} are the parabolic cylinder functions $D_{\nu_{\bf{k}}^*} (\pm(1+i)\tau)$ and their complex conjugates~\cite{grad}, where $\nu_{\bf{k}} = -1/2 + i\lambda_{\bf{k}}/2$. The particular solution depends on the initial condition chosen. If it is assumed that the field modes start out in the instantaneous vacuum state (vanishing particle number) in the asymptotic past ($\tau \to -\infty$), then the wave function of each oscillator must satisfy
\beq
\lim_{\tau \rightarrow -\infty} \psi_{\bf k}(\tau) \longrightarrow \left( \frac{\omega_{\bf k}(\tau)}{\pi}\right) ^{1/4} \exp \l(-\frac{\omega_{\bf k}(\tau)}{2}q_{\bf k}^2 - \frac{i}{2}\int^{\tau} \omega_{\bf k} d\tau \r)  
\eeq
which translates into the following condition on $\mu_{\bf k}$:
\beq
\lim_{\tau \to -\infty} \frac{\mu'_{\bf k}}{\mu_{\bf k}} ~\approx~ i\omega_{\bf k}(\tau)  
\label{limeq}
\eeq
where the prime denotes differentiation w.r.t. $\tau$. ($qE>0$ has been assumed, so $t$ and $\tau$ have the same sign in the asymptotic limits.) The particular solution of \eq{eom} which has the required behavior is $D_{\nu_{\bf{k}} ^{*}}[-(1+i)\tau]$. This may be seen by using the asymptotic form of this function in the $\tau \rightarrow -\infty$ limit:
\beq
\lim_{\tau \rightarrow -\infty} D_{\nu_{\bf{k}} ^{*}}[-(1+i)\tau] ~\approx~ (\sqrt{2}|\tau|)^{\nu_{\bf{k}} ^{*}} e^{\frac{i\pi}{4}\nu_{\bf{k}} ^{*}-\frac{i\tau^{2}}{2}}
\eeq
which gives
\beq
 \frac{\mu'_{\bf k}}{\mu_{\bf k}} ~\approx~ i|\tau| - \frac{\nu_{\bf{k}} ^{*}}{|\tau|} ~\approx~ i \omega_{\bf k}(\tau) -\frac{1}{2}\frac{\omega'_{\bf k}(\tau)}{\omega_{\bf k}(\tau)}   \label{f_early}
\eeq
and the $\omega'_{\bf k}/\omega_{\bf k}$ term vanishes in the $\tau \to -\infty$ limit. We next need to obtain the behavior at late times, for which the $\tau \rightarrow \infty$ limit of $\mu_{\bf k}$ is required. This is given by
\begin{eqnarray}
 D_{\nu_{\bf{k}} ^{*}}[-(1+i)\tau] &\stackrel{\tau \to \infty}{\longrightarrow}& e^{-i\pi\nu_{\bf{k}} ^{*}}(\sqrt{2}\tau)^{\nu_{\bf{k}} ^{*}} e^{\frac{i\pi}{4}\nu_{\bf{k}} ^{*}-i\frac{\tau^{2}}{2}} \l(1 + i\frac{\nu_{\bf{k}} ^* \l(\nu_{\bf{k}} ^* - 1 \r)}{4 \tau^2} \r) \nn \\
&~& + \frac{\sqrt{2\pi}}{\Gamma (-\nu_{\bf{k}} ^{*})} e^{-i\frac{\pi}{2}(\nu_{\bf{k}} ^{*}+1)}(\sqrt{2}\tau)^{\nu_{\bf{k}}} e^{-i\frac{\pi}{4}\nu_{\bf{k}} +i\frac{\tau^{2}}{2}} \l( 1 - i\frac{\l(\nu_{\bf{k}} ^* + 1 \r)\l(\nu_{\bf{k}} ^* + 2 \r)}{4 \tau^2}\r) \nn \\
&\equiv& {\cal A}_{\bf{k}} \frac{e^{-i\theta_{\bf{k}}(\tau)}}{\sqrt{\tau}}\l(1 + \frac{c_1}{\tau^2}\r) + {\cal B}_{\bf{k}} \frac{e^{i\theta_{\bf{k}}(\tau)}}{\sqrt{\tau}}\l(1 + \frac{c_2}{\tau^2}\r),
\end{eqnarray}
where
\beq
{\cal A}_{\bf{k}} = (\sqrt{2})^{\nu_{\bf{k}} ^{*}}e^{-i\frac{3\pi}{4}\nu_{\bf{k}} ^{*}} \qquad {\cal B}_{\bf{k}} = -i\frac{\sqrt{2\pi}}{\Gamma (-\nu_{\bf{k}} ^{*})}(\sqrt{2})^{\nu_{\bf{k}}}e^{-i\frac{\pi}{4}(2\nu_{\bf{k}} ^{*}+\nu_{\bf{k}})} \qquad \theta_{\bf{k}}(\tau) = \frac{\tau^{2}}{2} + \frac{\lambda_{\bf{k}}}{2}\ln(\tau)
\eeq
and
\beq
c_1 = i\frac{\nu_{\bf{k}} ^* \l(\nu_{\bf{k}} ^* - 1 \r)}{4} \quad , \quad c_2 =  - i\frac{\l(\nu_{\bf{k}} ^* + 1 \r)\l(\nu_{\bf{k}} ^* + 2 \r)}{4}.
\eeq
Using this approximation, it can be shown that in the late time limit,
\beq
z_{\bf k} ~\approx~ \frac{e^{-2i\theta_{\bf{k}}(\tau)}}{\cal R}_{\bf{k}} \l[ 1 - \frac{1}{4 \tau^2} \l(1+ \frac{e^{-2i\theta_{\bf{k}}(\tau)}}{\cal R}_{\bf{k}} \r)\l(i\l(1+{\cal R}_{\bf{k}} e^{2i\theta_{\bf{k}} (\tau)}\r) -2{\cal R}_{\bf{k}} e^{2i\theta_{\bf{k}} (\tau)}\l(c_1 - c_2\l(2-{\cal R}_{\bf{k}} e^{2i\theta_{\bf{k}}(\tau)}\r)\r)\r) + {\cal O}\l(\tau^{-3}\r) \r]   \label{z_E_adiab}
\eeq
where ${\cal R}_{\bf{k}} = {\cal B}_{\bf{k}}/{\cal A}_{\bf{k}}$. Using the above asymptotic form for $z_{\bf k}$, the late time limit of $\langle n_{\bf k}\rangle$ can be determined:
\beq
 \lim_{\tau \to \infty} \langle n_{\bf k}\rangle ~\equiv~ \lim_{\tau \to \infty}\frac{|z_{\bf k}|^2}{1-|z_{\bf k}|^2} ~=~ \frac{1}{(|{\cal R}_{\bf{k}}|^2-1)}. 
\eeq
Substituting the exact expressions for ${\cal A}_{\bf{k}}$ and ${\cal B}_{\bf{k}}$ gives 
\beq
\vert {\cal R}_{\bf{k}} \vert ^{2}=\frac{\vert {\cal B}_{\bf{k}} \vert ^{2}}{\vert {\cal A}_{\bf{k}} \vert ^{2}}=\frac{2\pi}{\vert \Gamma(-\nu_{\bf{k}} ^{*})\vert ^{2}} \vert e^{i\frac{\pi}{4}(\nu_{\bf{k}} ^{*}-\nu_{\bf{k}})} \vert ^{2}=2e^{\pi\lambda_{\bf{k}} /2}\cosh\l(\frac{\pi\lambda_{\bf{k}}}{2}\r) 
\eeq
and hence
\beq
 \lim_{\tau \to \infty} \langle n_{\bf k}\rangle ~=~ \exp\l(-\pi \lambda_{\bf{k}} \r) ~\equiv~ \exp\l(-\pi \frac{k_{n}^2+m^2}{qE} \r).\label{n_ef}
\eeq 
This expression is, of course, identical to the standard result which is normally computed using in-out states~\cite{schwinger,efield}. This agreement in the asymptotic limit shows that our choice of the time-dependent particle number to be a reasonable one. 

\begin{figure}[h]
\begin{center}
\subfigure[ ]{\label{efn}\includegraphics[width=8cm,angle=0.0]{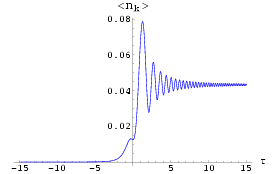}}
\hspace{0.6cm}
\subfigure[ ]{\label{efz}\includegraphics[scale=0.6]{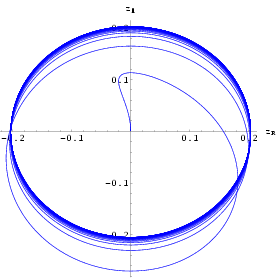}}
\end{center}
\caption{Plots of the particle number $\langle n_{\bf k} \rangle$ and the complex trajectory of $z_{\bf k}$ with time for the constant electric field case ($\omega_{\bf k}(\tau)=\sqrt{\lambda_{\bf k}+\tau^2}$) for $\lambda_{\bf k}=1$. $\langle n_{\bf k} \rangle$ saturates at a finite value at late times, $z_{\bf k}$ starts from zero, and at late times ends up going around in a circle centered on the origin of the complex plane.}
\label{ef}
\end{figure}

Given the exact solution, one can obtain the mean particle number as a function of time in terms of parabolic cylinder functions. The analytic forms are not illuminating but the plot giving 
the variation of the mean particle number as a function of the dimensionless variable $\tau$ --- given in figure~\ref{efn} for the choice of $\lambda_{\bf k}=1$ --- has several interesting features.   The particle number $\langle n_{\bf k} \rangle$ is nearly constant at very early and very late times as expected. It starts from zero in the asymptotic past, because of our specific choice of initial state. At late times, in the $\tau \to \infty$ limit, it  saturates at the constant value given by \eq{n_ef}. It may be noted that since the variable $\tau$ actually depends on both the original time coordinate $t$ and the momentum $k_z$, the value of $\tau$ for a given $t$ varies from mode to mode, implying that modes with different momenta would saturate and reach the adiabatic phase corresponding to $\tau \gg 1$ at different times. More importantly, at intermediate times (when the adiabaticity parameter $\epsilon$ is appreciably non-zero), $\langle n_{\bf k} \rangle$ is characterized by oscillatory behavior superimposed on a steadily increasing mean, which can be interpreted as a continual interplay of creation and annihilation of quanta. The amplitude of these oscillations gets progressively diminished as the evolution proceeds into the late time adiabatic regime, and the mean settles down at a constant value.  
(Though Schwinger effect is extensively discussed in the literature, the authors are not aware of a plot like the one in figure~\ref{efn} in any of the works.)

A handle on the nature of the time-evolution of the oscillator can be gained by directly visualizing the evolution of $z_{\bf k}$ in the complex plane. This is shown in figure~\ref{efz}. The complex trajectory of $z_{\bf k}$ starts from the origin, and after meandering around for a while in the intermediate phase, ultimately ends up, for large values of $\tau$, on a $|z_{\bf k}| = constant$ curve, corresponding to an asymptotically constant mean particle number. 

When we study particle production in cosmological backgrounds --- especially in the context of inflationary models --- it is usual to focus attention on the \emph{power spectrum}, which is related to the Fourier transform of the two-point correlation function for the field. To make latter comparison between Schwinger effect and cosmological particle production easier, we shall compute this for the Schwinger effect.
 The power in the ${\bf k}$th mode is just the variance of $q_{\bf k}$ in the given quantum state, and it can be written as
\beq
\langle q_{\bf k}^{2} \rangle = \frac{1}{2\l(R_{\bf k} + R^{*}_{\bf k}\r)} = \frac{\l ( 2 \langle n_{\bf k} \rangle + 1 \r)}{2 \omega_{\bf k}} + \frac{(\langle n_{\bf k} \rangle +1)}{ \omega_{\bf k} } \textrm{Re}(z_{\bf k}).  \label{ps_ef}
\eeq
The power, though related to the time dependent mean particle number, is not completely expressible in terms of it, being also dependent on the phase of $z_{\bf k}$ (the particle number however involves only the magnitude of $z_{\bf k}$). Only under the condition of $z_{\bf k}$ taking on a constant value, are the two quantities directly connected. In the present situation, $z_{\bf k}$ remains oscillatory in the adiabatic limit at late times while the particle number approaches a constant value, so the power, for large $\tau$, is approximately given by
\beq
\langle q_{\bf k}^{2} \rangle \approx \l(\frac{ 2 e^{-\pi \lambda_{\bf k}}  + 1 }{2 \tau}\r) + \l(\frac{e^{-\pi \lambda_{\bf k}} +1}{ \tau }\r) \textrm{Re} \l(\frac{e^{-i \l(\tau^{2} + \lambda_{\bf k}\ln \tau \r)}}{{\cal R}_{\bf{k}}} \r)  \label{psef}
\eeq 
which can be written as
\beq
\langle q_{\bf k}^{2} \rangle \approx \frac{1}{\tau}\l(e^{-\pi \lambda_{\bf k}}  + \frac{1}{2}+ \sqrt{e^{-\pi \lambda_{\bf k}}\l( e^{-\pi \lambda_{\bf k}} + 1 \r)} \cos \l( \tau^{2} + \lambda_{\bf k}\ln \tau + r_{\bf{k}} \r) \r)
\eeq
where ${\cal R}_{\bf{k}} = \vert {\cal R}_{\bf{k}} \vert \exp(ir_{\bf{k}})$. If the third term in the brackets, which is rapidly oscillating, is dropped, then  the variance at the leading order is just 
\beq
\langle q_{\bf k}^{2} \rangle \approx\frac{1}{\tau}\l( \langle n_{\bf k}\rangle + \frac{1}{2} \r)
\label{powSE}
\eeq
 and the power in the adiabatic limit can be related to the mean particle number. But this connection is not quite clear-cut for two reasons. First, it requires our ignoring the additional oscillatory term which allows us to ignore the phase information contained in $z_{\bf k}$. Second, the correlation actually dies down as $\tau$ when $\tau \to \infty$ and its interpretation is unclear.
 This point will be taken up again when making a comparison with the field theory in the de Sitter background.
 
\begin{figure}[h]
\subfigure[ ]{\includegraphics[width=6cm,angle=0.0]{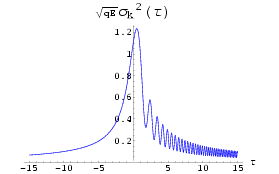}}
\subfigure[ ]{\includegraphics[width=6cm,angle=0.0]{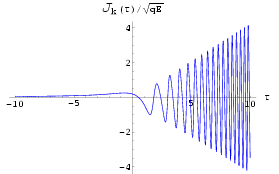}}
\caption{Variation of the functions $\sqrt{qE} \sigma_{\bf k} ^2$ and ${\cal J}_{\bf k}/\sqrt{qE}$ for the electric field case (for $\lambda_{\bf k}=1$). The Wigner function starts uncorrelated and peaked on the $p-$axis, and again at late times ends up peaking on the $p-$axis. }
\label{efW}
\end{figure}

Let us now turn to considering the question of quantum versus classical nature in terms of the phase space evolution of the Wigner function given by \eq{wigfun}, particularly in the asymptotic limits. 
The relevant functions which determine the Wigner function, $\sigma_{\bf k}^2$ and ${\cal J}_{\bf k}$ are shown in figure~\ref{efW}.

As is clear from these plots, the Wigner function starts uncorrelated and sharply peaked on the vertical $p_{\bf k}$ axis. At late  times it again ends up being  concentrated on the $p_{\bf k}$ axis. This behavior may be compared with that of the corresponding classical phase space trajectory, which, in either asymptotic limit ($|\tau|\gg 1$) has the approximate form 
\beq
q_c ~\sim~ \frac{1}{(\lambda_{\bf k} + \tau^{2})^{1/4}} \textrm{Re} \l[ {\cal C} e^{i \int^{\tau} \sqrt{\lambda_{\bf k}+\tau'^{2}} d \tau'} \r]~\approx~\frac{1}{\sqrt{|\tau|}}\l(1 - \frac{\lambda_{\bf k}}{4 \tau^2}\r) \textrm{Re} \l[ {\cal C} e^{i \l(\frac{\tau^{2}}{2} + \frac{\lambda_{\bf k}}{2} \ln|\tau| \r)} \r] 
\eeq
where ${\cal C}$ is an arbitrary complex constant, with the momentum being
\beq
p_c = \dot{q}_c \approx - \frac{q_c}{2 \tau}  + \l( \sqrt{|\tau|} + \frac{\lambda_{\bf k}}{4 |\tau|^{3/2}} \r) \textrm{Re} \l[ i~{\cal C} e^{ i \l(\frac{\tau^{2}}{2} + \frac{\lambda_{\bf k}}{2} \ln|\tau| \r) } \r]. 
\eeq
It is clear that in both asymptotic limits, $q_c \to 0$ and the trajectory is along the $p_{\bf k}$ axis. In other words the Wigner function is concentrated on the classical trajectory both at very early times (when the field is in the \emph{in} vacuum) and at late times (when particle number has reached an asymptotically constant limit. Hence, this aspect (i.e. peaking on the classical trajectory) of the evolutionary behavior of the Wigner function alone cannot be used in a direct manner to conclude the emergence of classicality. Obviously interpreting classicality merely in terms of peaking on the classical trajectory is suspect, since this can happen even when the oscillator is in a near-vacuum state (which happens to be the case here at early times). 

\begin{figure}[h]
\begin{center}
\includegraphics[width=9cm,angle=0.0]{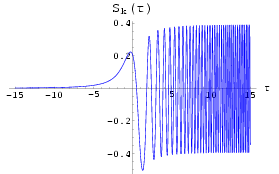}
\end{center}
\caption{Variation of the classicality parameter ${\cal S}_{\bf k}$ with time for the constant electric field case ($\lambda_{\bf k}=1$). ${\cal S}_{\bf k}$ is nearly zero in the asymptotic past, but ends up oscillating about zero with a finite amplitude at late times. The variation of ${\cal S}_{\bf k}$ reflects the complex trajectory of $z_{\bf k}$, which at late times is a circle centered on the origin of the complex plane.}
\label{efS}
\end{figure}

It would be of interest to compare the above behavior with that of the classicality parameter ${\cal S}_{\bf k}$ defined in \eq{classicality}, for this situation; its time variation is depicted in figure~\ref{efS}. This correlation measure clearly displays an asymmetry in time and more closely tracks the particle number: ${\cal S}_{\bf k}$ stays very close to zero in the course of the early time adiabatic evolution, but in the late time adiabatic regime, ends up oscillating about ${\cal S}_{\bf k}=0$. The oscillatory nature of ${\cal S}_{\bf k}$ reflects the oscillations of the imaginary part of $z_{\bf k}$, as it circles around; this becomes obvious by recalling the relation connecting $z_{\bf k}$ with ${\cal S}_{\bf k}$. As is evident from the plot, the time-averaged mean of this oscillatory variation is nearly zero, but the {\it variance} of ${\cal S}_{\bf k}$ has a non-zero finite value that stays nearly constant in the late time limit. Thus, although the Wigner function provides little help in addressing the question of classicality in this situation, the variation of ${\cal S}_{\bf k}$ points to the state becoming more correlated at late times, basically due to some amount of particle creation occurring over time, {\it in comparison} with the nearly zero value at early times. Nevertheless, none of these measures indicate a clear emergence of classicality in this context. We will see later that the situation is quite different in the case of evolution in the de Sitter background, showing a sharp difference between the Schwinger effect and evolution of a quantum field in a deSitter background.

One can also determine the form of the time-dependent effective Lagrangian $L_{eff}$ in the asymptotic limits from \eq{L_eff}. Using \eq{f_early}, it can be shown that $z_{\bf k}$ has the following approximate form in the adiabatic early time limit:
\beq
z_{\bf k} ~\approx~ -\frac{i}{4} \epsilon_{\bf k}(\tau) + {\cal O}\l(\epsilon^2\r)
\eeq
where $\epsilon_{\bf k}(\tau)=\omega'_{\bf k}/\omega^2_{\bf k}$, and the mean particle number is given by
\beq
\langle n_{\bf k}\rangle ~\approx~ \frac{\epsilon^2_{\bf k}(\tau)}{16} + {\cal O}\l(\epsilon^3\r).
\eeq
The asymptotic form of $z_{\bf k}$ allows one to derive an approximation for $L_{eff}$ in the $\tau \to -\infty$ limit. The real part can consequently be written as
\beq
\textrm{Re}L_{eff} ~=~ -\frac{i}{4} \frac{\omega'_{\bf k}}{\omega_{\bf k}} \textrm{Im} (z_{\bf k}) ~\approx~ \frac{\omega_{\bf k}(\tau)\epsilon^2_{\bf k}(\tau)}{16} ~\equiv~ \langle n_{\bf k}\rangle \omega_{\bf k}(\tau)
\eeq
to lowest order in $\epsilon$. This expression has the simple interpretation of being the energy drained from the background electric field due to particle production, $-\omega_{\bf k} \langle n_{\bf k} \rangle$. (Recall that the effective Lagrangian and effective Hamiltonian differ by a sign in this case.) Thus, to lowest non vanishing order in $\epsilon$, the real part of the effective Lagrangian incorporates information about particle creation during adiabatic evolution starting from a vacuum state, and this suggests that to lowest order in the adiabaticity parameter, the backreaction due to the production of particles can be incorporated using the first term in Re$L_{eff}$.

As for the late time adiabatic limit of large $\tau$, the asymptotic form of $z_{\bf k}$ in \eq{z_E_adiab} suggests that at the lowest order,  the real part of the effective action (per mode) will pick up dominant contribution over the late time adiabatic phase of evolution from the non-oscillatory term of magnitude $\epsilon^2_{\bf k}/16$ in the expression for $\textrm{Re} L_{eff}$. The imaginary part, on the other hand, is directly related to the asymptotic particle content with respect to the out vacuum state. One can further obtain an expression for the effective action for the field itself, by using the result that the contribution coming from one mode is just $(1/4)\ln(1+\langle n_{\bf k}\rangle)$ (see Eq.(38) of Paper I). We get:
\beq
\textrm{Im} A_{eff} = \frac{1}{4} \sum_{\bf{k}} \ln \l(1+e^{-\pi\lambda}\r) = \frac{V}{2}\int\frac{d^{3}{\bf{k}}}{(2\pi)^{3}}\sum_{n=1}^{\infty} (-1)^{n+1}\frac{e^{-\pi n \lambda}}{n}
\eeq 
To arrive at the last expression we have summed over every ${\bf k }$ twice, since each mode of a complex field is associated with two degrees of freedom, and used the Taylor series representation of $\ln(1+e^{-\pi\lambda})$. The summation over $\bf{k}$ has been transformed into an integral by the standard procedure of box normalization with periodic boundary conditions over a (large) volume $V$. The integrand has no dependence on the z-component of the momentum; so making the change $\int d^{3}{\bf{k}} \rightarrow \int dk_{z} \int 2\pi k_{n}dk_{n}$ and integrating over $k_{n}$ gives
\beq
\textrm{Im} A_{eff} = V \int \frac{dk_{z}}{qE}\sum_{n=1}^{\infty}\frac{(-1)^{n+1}}{n^{2}}\frac{(qE)^{2}}{(2\pi)^{3}}e^{-\frac{\pi nm^{2}}{qE}}
\eeq
Making the replacement $k_{z}/qE \to T$ (regarding $T$ as a time coordinate) brings the above expression into the form
\beq
\textrm{Im} A_{eff} = VT\sum_{n=1}^{\infty}\frac{(-1)^{n+1}}{n^{2}}\frac{(qE)^{2}}{(2\pi)^{3}}e^{-\frac{\pi nm^{2}}{qE}}
\eeq
which is of course the same as the standard expression obtained using in-out states~\cite{schwinger,efield}.

To summarize, our analysis of the time evolution in a constant electric field, which maps to a harmonic oscillator problem with  late time adiabatic evolution, reproduces all the standard results. In addition it indicates that the peaking of the Wigner distribution on the classical phase trajectory to be a fairly general feature, quite independent of the amount  particle production. It occurs even in the course of adiabatic evolution and in the initial quantum vacuum state. The $q$-$p$ correlation, in contrast, remains bounded and oscillatory in the late time adiabatic regime where particle production is suppressed, and is evidently more sensitive to the nature of the evolution, viz. whether it is adiabatic or non-adiabatic.  The relevance of these results will become more apparent when we study the evolution in cosmological background.

We shall now take up another scenario, that of expanding spacetime backgrounds, and consider time-evolution of a quantized field in specific models which illustrate both the kinds of late time evolution: adiabatic as well as non-adiabatic. 
 
\subsection{Quantum evolution in expanding universes}

The Friedmann model provides an instructive setting for analyzing quantum effects in a time-dependent gravitational field. A spatially flat Friedmann universe~\cite{txts1,infl1,infl2,paddy03,cosmo} is described by the line element
\beq
ds^2=dt^2-a^{2}(t)\, d{\bf x}^2
= a^{2}(\eta)\l(d\eta^{2} - d{\bf x}^2\r),\label{frwmet}
\eeq 
where $t$ is the cosmic time, $a(t)$ is the scale factor and $\eta$ denotes the conformal time with $d\eta= dt/a(t)$. We will consider a massless and minimally coupled real scalar field $\Phi$ in this spacetime (which, for example, could even describe linearized tensor perturbations \cite{txts1,infl1,infl2,cosmo}), governed by the covariant action
\beq
{\cal A}[\Phi] =\frac{1}{2}\int d^4x\, \sqrt{-g}\; 
\pa_{\mu}\Phi\, \pa^{\mu}\Phi.\label{actnPhi}
\eeq
The homogeneity and isotropy of the Friedmann metric~(\ref{frwmet}) allows one to decompose the scalar field $\Phi$ as
\beq
\Phi({\bf x}, \eta)
= \int\frac{d^{3}{\bf k}}{(2\pi)^{3/2}}\, 
Q_{\bf k}(\eta)\, e^{i{\bf k}\cdot{\bf x}}.
\label{Phi}
\eeq  
The $Q_{\bf k}$ are in general complex and can be written as $Q_{\bf k} = a_{\bf k} + i b_{\bf k}$, $a_{\bf k}$ and $b_{\bf k}$ being real. But since $\Phi$ is a real field, $Q_{-\bf k}=Q^{*}_{\bf k}$ and this imposes the constraints $a_{-\bf k}=a_{\bf k}$ and $b_{-\bf k}=-b_{\bf k}$, which means there actually is only one independent degree of freedom for every ${\bf k}$, and this can be chosen to be $a_{\bf k}$ for one half of $k$-space and $b_{-\bf k}$ for the other half. Relabeling these real variables as $q_{\bf k}$ over all of $k$-space, the action can then be expressed as
\beq
{\cal A}[\Phi]=\frac{1}{2}\int d^{3}{\bf k}\int d\eta\; a^2\,
\l(q_{\bf k}'^{2} -k^2\, q_{\bf k}^2\r),
\label{actnq}
\eeq
where the primes denote differentiation with respect to the conformal time $\eta$ and  $k=\vert {\bf k}\vert$. The action~(\ref{actnq}) describes a collection of independent oscillators with time dependent mass $m = a^2$ and frequencies $\omega = k$. The quantized field can thus be analyzed mode-by-mode, and the general formalism developed to study a time dependent oscillator can be directly applied here.

Let us now quickly recall some standard notions. In an expanding universe (described by a scale factor that is an increasing function of the cosmic time $t$), all physical length scales get stretched by the factor $a(t)$. Depending on the equation of state of the matter source that drives the expansion, the Friedmann model can undergo acceleration (characterized by $\ddot{a}(t)>0$) or deceleration ($\ddot{a}(t)<0$). The scale that determines the dynamics of a field, say, in such a background is set by the Hubble radius, defined as $R_{H}(t) = H(t)^{-1} = (\dot{a}/a)^{-1} = (a'/a^{2})^{-1}$. Any given field mode with a comoving wave vector $\bf{k}$ will have its physical wavelength red-shifted by the $a(t)$ factor, and there are two possible regimes: one, when the proper wavelength is smaller than $R_{H}$ (sub-Hubble scales) and the other when the mode is `outside' the Hubble radius, its physical scale exceeding $R_H$ (super-Hubble scales). The dynamics of the evolution of field modes in these two limits can be qualitatively quite different. 

The adiabaticity parameter for any given mode in this setting has the form $\epsilon(\eta)= 2 a'/ (a |{\bf k}|)$ (or in terms of cosmic time, $\epsilon(t)= 2 \dot{a}/ |{\bf k}|\propto \lambda(t)/R_H(t)$), which is essentially the ratio of its physical wavelength to the Hubble radius. This immediately tells us that in an accelerating universe (characterized by $\ddot{a}>0$), the adiabaticity parameter is an increasing function of time, so any given mode is expected to make a transition from initially adiabatic to highly non-adiabatic evolution at sufficiently large times, following Hubble exit. This trend will reverse in the case of decelerating models ($\ddot{a}<0$) and any mode will eventually end up evolving adiabatically. 

These features clearly show that  this scenario of a quantum field in an expanding gravitational background is quite different (conceptually and algebraically) from that  of the constant electric field case discussed earlier. In the case of electric field the evolution was adiabatic both at early and late times allowing us to define asymptotic vacuum states. Rigorously speaking, we can define an adiabatic in-vacuum at early times in the case of accelerating models while one can define an adiabatic out-vacuum at late times in the case of decelerating models.   In such  situations, the time-dependent definition of the particle concept is expected to provide a physically reasonable alternative, and will be adopted in our analysis of the quantized field in the examples which follow. 

On general grounds, particle creation effects are likely to be significant only in the non-adiabatic, i.e. super-Hubble phase of evolution (where $\epsilon_{\bf k}(\eta) \equiv k/aH \gtrsim 1 $). The expectation of strong late time particle production in inflationary universes, contrasted by saturation of the particle content as well as energy in the decelerating models at large times,  naturally follows from this. (We again stress that terms like `particle production' are used here in a specific sense, based on the idea of instantaneous vacuum, and should not be misinterpreted.) 

The standard approach to determine the evolution of the system starts with the equation for $\mu_{\bf k}$, \eq{mueq}. (Isotropy of the background means that all quantities will be functions of only $|{\bf k}|=k$, and this allows us to replace the label ${\bf k}$ by $k$; we shall stick to this notation from now on.) The solutions of \eq{mueq} are usually straightforward to obtain, and once the particular solution for a given initial condition has been chosen, it completely fixes the quantum state. The table below lists the various physical quantities expressed in terms of $z_k$ (which is directly computable from $\mu_k$):
\vspace{0.2cm}
\begin{center}
  {\scriptsize
\begin{tabular}{|c|c|}
  \hline
 & \\
  The wave function & $\psi (q_k,\eta) = \exp \l[ - \frac{k a^2}{2} \l(\frac{1-z_k}{1+z_k} \r) q_k^{2} \r] $\\
& \\
\hline
&  \\
Mean particle number & $\langle n_k \rangle = \frac{|z_k|^2}{1-|z_k|^2}$   \\
&  \\
\hline
&  \\
Mean energy per mode & ${\cal E}_k = \l( \langle n_k \rangle + \frac{1}{2} \r) k$   \\
&  \\
\hline
&  \\
 Wigner function ${\cal W}\l(q_k, p_k, t\r)$ &  $\sigma_k ^2 = \frac{|1+z_k|^2}{k a^2 (1-|z_k|^2)}$  \\
 &  \\
~~~$=\frac{1}{\pi}\; 
\exp\l[-\frac{q_k^2}{\sigma_k^2(t)}
-\sigma_k ^2(t)\, \l(p_k-
{\cal J}_k(t)\,  q_k\r)^2\r] $ &~~~${\cal J}_k = \frac{2 k a^2 Im(z_k)}{|1+z_k|^2}$ \\
& \\
\hline
& \\
Classicality parameter &   $ {\cal J}_k \sigma_k ^{2}  = 2 \langle p_k q_k \rangle_{{\cal W}} $ \\
& \\
${\cal S}_k = \frac{{\cal J}_k \sigma_k ^2 }{\sqrt{1 + ({\cal J}_k \sigma_k ^2 )^{2}} }$  &  $= \frac{2 Im(z_k)}{(1-|z_k|^{2})}$\\
& \\
\hline
& \\
The power spectrum &  $\langle q_k^{2} \rangle = \frac{\langle n_k \rangle}{2 k a^2}  \l| 1+\frac{1}{z_k} \r| ^{2}$ \\
& \\
$k^3 P_{\Phi} (k) = \frac{k^3}{2 \pi^2} \langle q_k^{2} \rangle$  &  = $\frac{\l ( 2 \langle n_k \rangle + 1 \r)}{2 k a^2} + \frac{(\langle n_k \rangle +1)}{ k a^2 }Re (z_k)$ \\
& \\
\hline
& \\
& $\textrm{Re} L_{eff} = -\frac{1}{4} \frac{m'_k}{m_k} Im (z_k)$ \\
Effective Lagrangian  &  \\ 
& ~~$\textrm{Im} L_{eff} = \frac{1}{4} \frac{m'_k}{m_k} Re (z_k) = \frac{1}{4} \frac{d}{d \eta}\ln(1+\langle n_k \rangle)$~~ \\
& \\
\hline

\end{tabular}
}
\end{center}
\vspace{0.3cm}

Our basic interest  is in  understanding how the time-dependent particle content and the Wigner function describing the phase space correlations of the quantum state  evolve in the expanding spacetime background in various limits. To keep the analysis clear and simple, we shall consider cosmological models characterized by power-law expansion with the scale factor having the form $a(t) \propto t^p$, $p$ lying in the range $[0,\infty]$. This set covers accelerating ($p>1$) as well as decelerating ($0<p<1$) types. As mentioned before, our intention is to consider examples that illustrate the two possible extreme cases of late time evolution, and the power law models prove adequate for this purpose. In particular, the accelerating models have close correspondence with the toy model of the frequency function $\omega_3$ we considered earlier, which is non-adiabatic at large times, while the non-inflationary ones share the feature of an asymptotically adiabatic phase in common with two other toy models (those with the frequency functions $\omega_1$ and $\omega_2$). We shall begin by picking one representative example from each of these two distinct groups, the de Sitter model and the radiation-dominated model respectively, and try to bring out the essential features that distinguish (or are shared in common by) them. Following this, a model combining an initial decelerating phase with a late time accelerated period of expansion will be analyzed, and finally a general approximate asymptotic analysis will be carried out separately for inflationary and decelerating models. 

\subsection*{1. The de Sitter universe}   \label{sec:ds}

The de Sitter universe~\cite{infl1,infl2,paddy03,cosmo} has exponential $a(t)$ and can describe the inflationary universe. It essentially corresponds to taking the limit of the index $p\to\infty$ in the power law models with suitable normalization.  In terms of the conformal time $\eta$, the scale factor is given by
\beq
a(\eta) = -\frac{1}{H \eta}
\label{dsa}
\eeq
where $\eta \leq 0$. The adiabatic parameter $\epsilon_k = 2/k|\eta| \equiv 2aH/k$ clearly blows up as the scale factor goes to infinity. This naturally means that every mode,  starting out at sub-Hubble scales at early times, will `exit' the Hubble radius and continue to remain super-Hubble thereafter. As indicated earlier, the divergence of $\epsilon_k(\eta)$  indicates  the absence of an adiabatic out-vacuum state, and hence a natural definition of particles at late times. Consequently, and definition for particle is intrinsically ad hoc and one need to base the discussion on something which is useful. We will see that the time-dependent instantaneous particle concept is a fairly reasonable choice in such a situation. 

\begin{figure}[h]
\begin{center}
\subfigure[ ]{\label{ni}\includegraphics[width=7cm,angle=0.0]{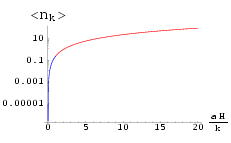}}
\hspace{0.7cm}
\subfigure[ ]{\label{zi}\includegraphics[scale=0.7]{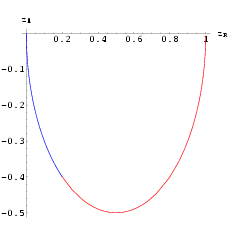}}
\end{center}
\caption{Plots of the particle number $\langle n_k \rangle$ and the complex trajectory of $z_k$ with the dimensionless variable $aH/k$ for the Bunch-Davies state in the de Sitter model. Sub- and super-Hubble regimes are color-coded blue and red respectively. $\langle n_k \rangle$ starts from zero and diverges at late times. The trajectory of $z_k$, starting from the origin, remains confined to a quadrant in the complex plane, and at late times (super-Hubble scales) approaches the point $z_k=1$.}
\label{inf}
\end{figure}

To determine the time evolution of the quantum state, we will begin with \eq{mueq}. For the expansion given by \eq{dsa} the general solution for $\mu_k$ can be expressed in terms of simple functions~\cite{grad,arf} as follows: 
\beq
\mu_{k}(a)
=  \frac{1}{\sqrt{2k} a}
\biggl[\l(1 - \frac{i a H}{k}\r)\, e^{-\frac{ik}{aH}}
+\; {\cal R}_k\, \l(1 + \frac{i a H}{k}\r)\, e^{\frac{ik}{aH}}\biggr]
\label{eq:gsei}
\eeq
where ${\cal R}_k$ is an arbitrary complex number. If one assumes the mode to be set in the instantaneous ground state at the moment $a=a_i$, this corresponds to imposing the condition $R_k(a_i)  = k\, a_{i}^2/2$. The function ${\cal R}_k$ then turns out to be
\beq
{\cal R}_k=\l(1- \, \frac{2 i k}{a_i H}\r)^{-1}\, \exp \l(-\frac{2 i k}{a_i H}\r).
\label{eq:BkD}
\eeq
Here, we shall choose to work with a state that begins in an instantaneous vacuum in the asymptotic past; this implies $a_i\to 0$, so that ${\cal R}_k$ vanishes for all $k$. This state is generally known in the literature as the Bunch-Davies vacuum~\cite{txts1,infl1,infl2}, and its evolution will be considered in the analysis which follows. 
Once the form of $\mu_k$ is specified, it is trivial to show that 
\beq
z_k(a) = \frac{i}{i - \frac{2k}{aH}} \equiv \frac{1}{1 + \frac{2ik}{aH}}
\eeq
which is quite simple. From this it follows that the time-dependent mean particle number is given by the following expression:
\beq
\langle n_{k} \rangle = \frac{|z_k \l( a \r)|^2}{1-|z_k \l( a \r)|^2} = \frac{a^2 H^2}{4 k^2 } \equiv \frac{1}{4}\l(\frac{\lambda_p}{R_H}\r)^2 
\eeq
$\lambda_p$ being the physical wavelength of the mode. The particle number sharply diverges in the late time non-adiabatic regime, i.e. at super-Hubble length scales ($k/aH \ll 1$) as the accelerated expansion drives rapid particle production. The time-variation of $\langle n_{k} \rangle$ as well as the trajectory of $z_k$ in the complex plane for the Bunch-Davies state are depicted in figures~\ref{ni} and \ref{zi}. $z_k$ starts from the origin, and eventually ends up at the limiting point corresponding to $z_k=+1$. This is in sharp contrast to the evolution in the case of Schwinger effect, in which the $z$ goes around a closed loop in the complex plane at late times (see figure~\ref{efz}).

It may be noted that since the mean particle number here grows monotonically with time, one can trade off the time dependence for dependence on $\langle n_{k} \rangle$, and therefore express $z_k$ entirely in terms of the mean number of particles:
\beq
z_k(\langle n_{k} \rangle) ~=~ \frac{1}{1 + \frac{2ik}{aH}} ~=~ \frac{\sqrt{\langle n_{k} \rangle}}{i + \sqrt{\langle n_{k} \rangle}}.
\eeq
This in turn allows one to express the wave function, too, explicitly as a function of just the mean particle number:
\beq
\psi_k (q_k, \langle n_{k} \rangle) ~=~ \l[ \frac{4 k^3}{\pi H^2} \l(\frac{\langle n_{k} \rangle}{1 + 4 \langle n_{k} \rangle}\r) \r]^{1/4} \exp \l[-\frac{2 k^3}{H^2}\l( \frac{\langle n_{k} \rangle}{1 - 2 i \sqrt{\langle n_{k} \rangle}}\r) q_{k}^2 \r].
\eeq
This is possible here only because of a one-to-one relation between $a$ and $\langle n_{k} \rangle$. Such a transformation will not be unique in general (for e.g. if the mean particle number is oscillatory, the inversion would give multiple values of time for the same $\langle n_{k} \rangle$).
   
In the literature, what is normally computed for inflationary models is not the particle number but the power spectrum of the field~\cite{infl1,infl2,ps,paddy03}, and we would like to connect it up with the particle concept used above. As mentioned earlier, the power in the ${\bf k}$th Fourier mode is just the variance of the $q_k$ variable in the given quantum state, and is expressible as follows:
\beq
\langle q_k^{2} \rangle = \frac{1}{2\l(R_k + R_k^*\r)} = \frac{\l ( 2 \langle n_k \rangle + 1 \r)}{2 k a^2} + \frac{(\langle n_k \rangle +1)}{ k a^2 }\textrm{Re}(z_k).  \label{ps}
\eeq
From \eq{ps}, it is clear that the power spectrum is not in one-to-one correspondence with the particle content, since it also depends on the  phase of $z_k$ through the Re($z_k$) term. It is only under certain conditions --- like when $z_k$ goes to a constant value --- that a direct correspondence between the two can be established. This is possible, for example, if the power is evaluated at super-Hubble scales ($k/aH \ll 1$), because in this limit, $z_k \to 1$, and from this it also follows that the particle number is large ($\langle n_k \rangle \gg 1$). Consequently, we have
\beq
\langle q_k^{2} \rangle \approx 2 \langle n_k \rangle /(k a^2)
\label{powdS}
\eeq 
and thus, at large scales, the power in any given mode does provide a measure of its particle content. 
Comparing \eq{powdS} with the corresponding equation \eq{powSE} in the case of Schwinger effect, we find that there are significant differences. First, the situation
is mathematically much clearer in the case de Sitter universe in which the asymptotic limit of $z$ is unambiguous. This is not the situation in the case of Schwinger effect and one need to drop oscillatory terms in a somewhat arbitrary fashion to arrive at \eq{powSE}. This can be tracked down to the fact that in the case of de Sitter universe the late time evolution is highly non-adiabatic while in the case of Schwinger effect the late time evolution is adiabatic. As shown in Paper I, the adiabatic evolution generally leads to a late time oscillatory behavior for $z_{\bf k}$ making the relation between power spectrum and particle content ambiguous.
 
For the case of de Sitter evolution starting from the Bunch-Davies state, the power is given by
\beq
\langle q_k^{2} \rangle = \frac{H^2}{2 k^3}\l( 1 + \frac{k^2}{a^2 H^2} \r). 
\eeq 
When evaluated for every mode at Hubble exit ($k/aH = {\cal O}(1)$), this expression yields $k^3 \langle q_k^{2} \rangle \propto H^2$, corresponding to a scale-invariant spectrum.  

\begin{figure}[h]
\subfigure[ ]{\label{sigi}\includegraphics[width=7cm,angle=0.0]{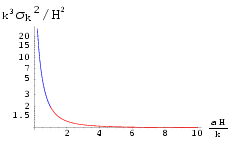}}
\hspace{0.5cm}
\subfigure[ ]{\label{ji}\includegraphics[width=7cm,angle=0.0]{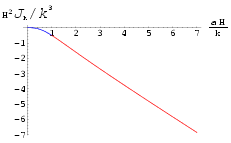}}
\caption{Variation of the dimensionless functions $k^3\sigma_k ^2/H^2$ and $H^2{\cal J}_k /k^3$ for the Bunch-Davies state. Sub- and super-Hubble regimes are color-coded blue and red respectively. The Wigner function starts uncorrelated and concentrated on the $q$-axis; at late times ($aH/k \to \infty$), $q/p \to 0$ representing a state concentrated on the $p$-axis. }
\label{Wi}
\end{figure}

We now turn our attention to understanding  the quantum versus classical nature of the state. We will start with the  evolution of the  Wigner function in phase space. The  Wigner function is determined in terms of the two variables $\sigma_k^2$ and ${\cal J}_{k}$ which are given by
\beq
\sigma_k^2(a) = \frac{H^2}{k^3}\l( 1 + \frac{k^2}{a^2 H^2} \r) ;\quad {\cal J}_{k}(a) = -\frac{a k^2}{H}\l( 1 + \frac{k^2}{a^2 H^2} \r)^{-1}.    \label{sig_j_i}
\eeq
Figures~\ref{sigi},\ref{ji} show the variation of these functions with time. From the above expressions, it is clear that ${\cal J}_{k}\to 0, \sigma_k^2\to\infty$ at early times ($k/aH \to \infty$) i.e. sub-Hubble scales, which corresponds to a state that starts sharply concentrated on the $q$-axis. At late times ($k/aH \to 0$), however, ${\cal J}_{k}\to \infty$ with a finite $\sigma_k^2$, which causes the Wigner function to end up peaking on the $p$-axis.

Let us compare the above behavior with that of the corresponding classical phase space trajectory. For the case of exponential inflation, the general solution for $q_{k}$ in terms of the time coordinate $\eta$ can be written as
\beq
q_{ k}= \textrm{Re}\, \left[
 {\cal C}   \eta\left(
1+\frac{i}{k\eta}\right) \, e^{ik\eta}
\right]
\eeq
where ${\cal C}$ is an arbitrary complex number. Writing 
 ${\cal C}=C\, e^{i\, c}$, this solution can be re-expressed as
\beq
q_{ k}= C\,   \eta\, \cos\l(k\eta+c\r)
- \frac{C\,   }{k}\, \sin\l(k\eta+c\r).    \label{q_inf}
\eeq
The conjugate momentum $p_{ k}=\l(a^2\, q_{ k}'\r)$ 
corresponding to the above $q_{k}$ is then given by
\beq
p_{ k}= -\l(C\, k/ H^2 \eta\r)\, \sin\l(k\eta+c\r).
\eeq
The classical trajectory of the system in the phase space thus takes the form
\beq
q_{ k} = \frac{H^2\eta}{k^2} p_{ k} \pm C\eta \left(1-\frac{H^4\eta^2}{C^2k^2}p_{ k}^2\right)^{1/2}
\label{classtraj1}
\eeq
where $\eta$ is to be understood as being an implicit function of $q_{k}$ (which can be found by inverting \eq{q_inf}). At late times ($\eta \to 0$) i.e. at super-Hubble scales, $(q_{ k}/p_{ k}) \to 0$, indicating a trajectory that ends up along the vertical $p_k$ axis, and this is precisely what the Wigner function gives. As for the other limit of $\eta \to -\infty$, we get $p_{ k}\to0$ with $\eta p_{k}$ remaining finite, which obviously gives a trajectory along the $q_k$ axis. This again matches with the analysis based on the Wigner function.
 Thus, the Wigner function is found to peak on the classical trajectory not only in the non-adiabatic late time phase when strong particle creation takes place, but even at very early times when the quantum state of the mode is in fact assumed to be close to a vacuum. This may be taken to suggest that this aspect of the evolutionary behavior of the Wigner distribution is not sufficient as far as providing information about classicality is concerned. We reached a similar conclusion in the case of the Schwinger effect.

In contrast, the classicality parameter gives a more meaningful description here. It can be directly determined from \eq{sig_j_i}, and is given by
\beq
{\cal S}_k (a) ~=~ - \frac{\l(a H/k \r)}{\sqrt{1+\l(a H /k \r)^2}} ~\equiv~ -2\sqrt{\frac{\langle n_k \rangle}{1 + 4\langle n_k \rangle}}.
\eeq
Let us look at the asymptotic limits of this expression. At sub-Hubble scales ($k/aH \to \infty$), the mean particle number is small, and
\beq
{\cal S}_k ~\approx~ - 2 \sqrt{\langle n_k \rangle} \l[ 1 + {\cal O}\l(\langle n_k \rangle \r) \r]
\eeq
while at late times, as $\langle n_k \rangle$ grows sharply at super-Hubble scales ($k/aH \to 0$),
\beq
{\cal S}_k ~\approx~ -1 + \frac{1}{8 \langle n_k \rangle}
\eeq
to the lowest order. (At Hubble crossing, $k/aH \sim 1$ and so ${\cal S}_k \sim 1/\sqrt 2$.) 
It is clear that this quantity ${\cal S}_k$ follows more closely the particle content, in that it remains close to zero in the initial adiabatic sub-Hubble phase, but grows subsequently, saturating at unity in the large $\langle n_k \rangle$ limit. (This, it may be remembered, was also found to happen in the non-adiabatic toy model corresponding to $\omega_3$ considered in Paper I~\cite{gm}.) Thus, if ${\cal S}_k$ is used as a measure of the degree of correlation, then one can unambiguously conclude that  the state is evolving towards classicality, which agrees with the standard lore in this subject. But note that the Wigner function peaks on the classical trajectory both at early and late times and hence is not such a good descriptor. 

\begin{figure}[h]
\begin{center}
\includegraphics[width=8cm,angle=0.0]{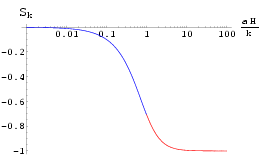}
\end{center}
\caption{Plot of the classicality parameter ${\cal S}_k$ for the Bunch-Davies state in the de Sitter model. Sub and super Hubble regimes are color-coded blue and red respectively. $\vert {\cal S}_k \vert$ starts from zero in the asymptotic past and saturates at unity at super-Hubble scales.}
\label{infS}
\end{figure}

The plot of ${\cal S}_k$ for the Bunch-Davies state is shown in figure \ref{infS}.

The contribution to the time-dependent effective Lagrangian coming from the ${\bf k}$th mode also can be computed using the expression for $z_k$, and is given by
\beq
L_{eff} = \frac{i}{4}\frac{m_k '}{m_k} z_k = \frac{k}{1 + \frac{4 k^2}{a^2 H^2}} \l( 1 + \frac{i a H }{2 k} \r)
\eeq
where the real part can be rewritten in terms of the mean number of particles in the mode {\bf k}, as
\beq
\textrm{Re} L_{eff} = \frac{k}{1 + \frac{4 k^2}{a^2 H^2} } \equiv \frac{k \langle n_k \rangle}{1 + \langle n_k \rangle}.
\eeq
In the asymptotic past ($k/aH \to \infty$), the imaginary part goes to zero, while the real part gives
\beq
\textrm{Re} L_{eff} ~\approx~  k \langle n_k \rangle + {\cal O}\l(\langle n_k \rangle^2 \r)
\eeq
(recalling that the mean particle number in the Bunch-Davies state vanishes in this limit). To lowest order in $\langle n_k \rangle$, this can be interpreted as the energy drained from the gravitational field that goes into creating particles, $-\langle n_k \rangle k$. This again suggests, as in the electric field case, that information about particle production during adiabatic evolution starting from a vacuum state (for the quantum field) is encoded in the real part of the effective Lagrangian, and that one can take into account backreaction due to the produced particles using Re$L_{eff}$. This makes sense and lends further credence to our time dependent definitions. 

In the other limit of late times (corresponding to $k/aH \to 0$), the imaginary part blows as a consequence of strong particle creation, while the real part assumes the approximate form
\beq
\textrm{Re} L_{eff} \approx  k \l( 1 - \frac{1}{\langle n_k \rangle} \r)
\eeq
which saturates at the limiting value Re$L_{eff} = k$ asymptotically. Clearly, in this limit where $\langle n_k \rangle \gg 1$, the correspondence of Re$L_{eff}$ with particle creation which holds in the adiabatic regime is not valid, and the interpretation of the above expression is presently unclear. 

Finally the physical content of the quantum state can also be clarified in terms of the spatial Fourier transform of $z_k$, which, as was shown earlier, can be thought of as the amplitude for a pair of particles to be produced from the vacuum at any given instant. This function $Z({\bf r},a)$ is given by    
\beq
Z({\bf r},a) = \int e^{i {\bf k \cdot r}} z_k \frac{d^3 {\bf k}}{\l( 2 \pi \r) ^3} = -\frac{1}{2 \pi^2 } \int_{0}^{\infty} \frac{k}{1+\frac{2ik}{aH}}\frac{\sin (kr)}{r} dk \label{ft_z}
\eeq 
which can be written in terms of standard functions:
\beq
Z(r,a) = \frac{i}{4 \pi^2} \l(\frac{a^2 H^2}{r}\r) \l[\frac{1}{r a H} + \frac{1}{4}\l( e^{\frac{r a H}{2}} Ei\l(-\frac{r a H}{2}\r) - e^{-\frac{r a H}{2}} Ei\l(\frac{r a H}{2}\r) \r)\r]
\eeq
$Ei(x)$ being the exponential integral function~\cite{grad}. While this is not very illuminating, it has clear asymptotic forms which confirms our interpretation.
When $a \to 0$, the second term in the square brackets above vanishes and
\beq
\lim_{a \to 0} Z(r,a) \sim \frac{i}{4 \pi^2} \l(\frac{a H}{r^2}\r).
\eeq
On the other hand, when $a \to \infty$ it reduces to a Dirac delta function:  
\beq
\lim_{a \to \infty} Z({\bf r},a) = \delta ({\bf r}).
\eeq 
(While one can get this result by a careful limiting procedure, it is obvious from the fact that at late times, $z_k \to 1$.) The first limit found above can be interpreted as suggesting that at early times, the particles created at two points are strongly correlated, implying an underlying quantum nature, and the correlation falls off with separation only as $r^{-2}$. The large $a$ limit of a delta function, on the other hand, implies particles with non-zero separation are uncorrelated, in line with the classical notion of a particle as a spatially localized entity. Thus, $Z(r,a)$ provides a measure of the classicality and particle content of the state consistent with what we found earlier.

To recapitulate, the de Sitter background provides an example in which the late time evolutions is highly non-adiabatic with no sensible out-vacuum in the conventional terminology. Our formalism, nevertheless produces sensible results --- consistent with what have been known in the literature --- and shows that the state evolves towards classicality. This is clear in the behavior of the classicality parameter (which increases from zero to unity in the course of evolution) and that of $Z({\bf r},a)$ which becomes a delta function at late times. The Wigner function, on the other hand, is not a good descriptor of classicality since it is peaked on the classical trajectory both at early and late times and does not track the classical trajectory in the intermediate epochs. This shows a contrasting behavior vis-a-vis the evolution in a constant electric field background which leads to adiabatic evolution both at early and late times allowing for adiabatic in-vacuum and out-vacuum to be defined. The evolution of the system towards classicality is much less clear in this case. Altogether, these results closely resemble those found for the non-adiabatic and adiabatic toy examples of Paper I~\cite{gm}. 

We now move on and choose an example of a decelerating universe, to illustrate the other alternative of late time adiabatic evolution. We expect this to  present a contrast with the behavior found above and resemble the Schwinger effect more closely.         

\subsection*{2. The radiation-dominated model}  \label{sec:rd}

We shall now work in the setting of radiation dominated expansion~\cite{paddy03,cosmo}, that corresponds to setting $p=1/2$. The radiation dominated model is thus governed by the expansion law
\beq
a(\eta) = a_{0} \eta \qquad (0 \leq \eta < \infty).   \label{rada}
\eeq
Here, the adiabaticity parameter given by $\epsilon(\eta) = 2/k\eta \equiv 2 a_0/(ka)$ steadily drops to zero with expansion, and particle production in any given mode is naturally expected to be strongly suppressed in the late time sub-Hubble phase of evolution. In fact, in being characterized by a late time adiabatic regime, the radiation dominated model bears close resemblance to the constant electric field case that was considered earlier and contrasts with the de Sitter example.

For the linear scale factor in \eq{rada}, $a''=0$ and the solutions of the classical mode equation \eq{mueq} are particularly simple, being just plane waves; the general solution for $\mu_k$ can be written as
\beq
\mu_{k}(a) = \frac{1}{\sqrt{2k} a}\l( e^{ik\frac{a}{a_0}} + {\cal R}_k e^{-ik\frac{a}{a_0}}\r)  \label{rmfm}
\eeq
${\cal R}_k$ being an arbitrary constant coefficient. If it is assumed that the mode starts off in the instantaneous vacuum state at the moment $a=a_i$, then ${\cal R}_k= - \l(1+2i\, ka_i /a_0\r)^{-1}\, e^{2 i k a_i/a_0}$, and it is straightforward to determine the form of $z_k$:
\beq
z_k(a) = \frac{2ik\l(\frac{a}{a_0}\r) {\cal R}_k e^{-2ik\frac{a}{a_0}}+\l(1+{\cal R}_k e^{-2ik\frac{a}{a_0}}\r)}{2ik\l(\frac{a}{a_0}\r)  - \l(1+{\cal R}_k e^{-2ik\frac{a}{a_0}}\r)}.
\eeq
If the state is extended into the past to $a<a_i$, $z_k \to -1$ and the particle number is found to diverge as the initial singularity is approached, for arbitrary ${\cal R}_k$, and it is not possible to define an instantaneous vacuum state at $a_i = 0$ when the spacetime is singular. Therefore, we have to confine ourselves to states which  start off in the ground state at some $a_i > 0$. The behavior of the field mode prior to $a_i$ will not be of relevance to us, and we will limit ourselves only to the subsequent evolution of the wave function (for $a>a_0$). This situation does not occur in the de-Sitter model, where one is free to choose the initial conditions at arbitrarily early times, like in the case of the Bunch-Davies state for which $a_i \to 0$.

\begin{figure}[h]
\begin{center}
\subfigure[ ]{\label{nr}\includegraphics[width=8cm,angle=0.0]{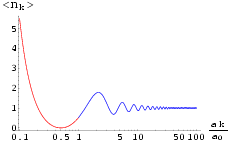}}
\hspace{0.4cm}
\subfigure[ ]{\label{zr}\includegraphics[scale=0.75]{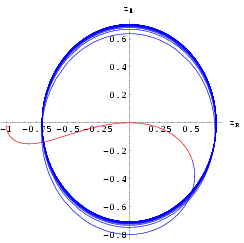}}
\end{center}
\caption{Plots of the particle number $\langle n_k \rangle$ and the complex trajectory of $z_k$ as functions of the dimensionless variable $ka/a_{0}$ for the radiation-dominated model, for the case when the initial condition is imposed at $ka/a_0 = 1/2$. Sub- and super-Hubble regimes are color-coded blue and red respectively. $\langle n_k \rangle$ saturates at a finite value at late times (sub-Hubble scales). $z_k$ starts from zero, and at late times ends up circling around the origin of the complex plane.}
\label{rad}
\end{figure}

We are particularly interested in the behavior of the quantum state at large times ($ka/a_0 \gg 1$). In this limit, $z_k$ assumes the approximate form
\beq
z_k(a) ~\approx~ {\cal R}_k e^{-2 i k \frac{a}{a_0}} - \frac{i}{2}\frac{a_0}{k a} \l( 1 + {\cal R}_k e^{-2 i k \frac{a}{a_0}} \r)^2 + {\cal O}\l( \frac{a_{0}^2}{k^2 a^2} \r).   \label{zr_adiab}
\eeq 
The above expression makes it  clear that in the adiabatic late time limit, the mean particle number in any given mode saturates at a finite value:
\beq
\lim_{ka/a_0 \to \infty} \langle n_{k} \rangle = \frac{|{\cal R}_k|^2}{1-|{\cal R}_k|^2} \equiv \frac{a_{0}^2}{4 k ^2 a_{i} ^2}.
\eeq
The plot of $\langle n_{k} \rangle$ (for the particular case where the instantaneous vacuum is chosen at $ka/a_0=1/2$) is shown in figure~\ref{nr}. Starting from zero, the mean number is characterized by transient oscillations that get steadily suppressed as the mode turns adiabatic. Figure~\ref{zr} displays the corresponding complex trajectory of $z_k$ which ends up going around in a circle centered on the origin of the complex plane, indicative of a finite, asymptotically constant particle number. 
These are identical in form to what we found in the case of the constant electric field.

\begin{figure}[h]
\subfigure[ ]{\label{sigr}\includegraphics[width=7cm,angle=0.0]{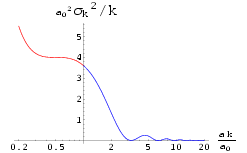}}
\hspace{0.5cm}
\subfigure[ ]{\label{jr}\includegraphics[width=7cm,angle=0.0]{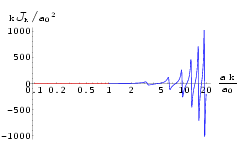}}
\caption{Variation of the dimensionless functions $a_{0} ^2 \sigma_k /k$ and $k{\cal J}_k /a_{0}^2$ for the radiation-dominated model. Sub and super Hubble regimes are color-coded blue and red respectively. The Wigner function starts uncorrelated at $k a/a_0 = 1/2$ with a finite value of $\sigma_k ^2$; at sub-Hubble scales, $\sigma_k ^2 \to 0$ causing the Wigner function to get peaked on the $p$-axis. }
\label{o1W}
\end{figure}

Let us look at the Wigner distribution in this case. It is again described in terms of the following two functions:
\beq
\sigma_k ^2(a) = \frac{\l|1 + {\cal R}_k e^{-2ik\frac{a}{a_0}}\r|^2}{\l(1-|{\cal R}_k|^2\r) k a^2}  \label{sig_r}
\eeq
and
\beq
{\cal J}_{k}(a) = \l[ \frac{2 k a^2 \textrm{Im}\l({\cal R}_k e^{-2ik\frac{a}{a_0}}\r)}{\l|1 + {\cal R}_k e^{-2ik\frac{a}{a_0}}\r|^2}- a_0 a \r],   \label{j_r}
\eeq
and these are depicted in figures~\ref{sigr} and \ref{jr}. The Wigner function clearly starts off in an uncorrelated state (${\cal J}_k =0$) at $a = a_i$ and with a finite spread of magnitude $\sigma_k^2=k^{-1}a_{i} ^{-2}$. However, at late times, $\sigma _{k} ^2 \to 0$ and this corresponds to a distribution sharply peaked along the $p$-axis. (This is presumably a consequence of the $a\to\infty$ limit of the scale factor.) 
As for the corresponding general classical phase space trajectory, it is given, in terms of the conformal time $\eta$ by
\beq
q_{ k} = \textrm{Re} \l( {\cal C}\frac{e^{ik\eta}}{\eta}\r) = C\frac{\cos(k\eta+c)}{\eta}  \label{q_rad}
\eeq
where ${\cal C}=C e^{ic}$ is an arbitrary complex number. The corresponding momentum has the form 
\beq
p_{ k}=a^2 q'_{ k} = -a_{0} ^2 C \l[ k\eta \sin(k\eta + c) + \cos(k\eta + c) \r].
\eeq
Expressing $q_{ k}$ in terms of $p_{ k}$ yields the following expression for a general classical trajectory:
\beq
q_{ k}= -\frac{p_{ k}}{a_{0} ^2 \eta (1+k^2\eta^2)} \pm \frac{C k}{\sqrt{1+k^2\eta^2}}\sqrt{1-\frac{p_{ k}^2}{a_0 ^{4} C^{2} (1+k^2\eta^2)}}
\label{classtraj2}
\eeq
where, again, $\eta$ is to be treated as an implicit function of $q_{ k}$, making use of \eq{q_rad}. Every trajectory at the initial moment $a_i$ starts from some point ($q_{k}(a_i),p_{k}(a_i)$) in the phase space. But we are particularly interested in the late time behavior here. It is evident that in this limit, $q_k \to 0$ and we get a trajectory along the $p$-axis. This matches with the corresponding limiting form of the Wigner function. Clearly, we here have yet another instance of the Wigner function showing peaking on the classical trajectory despite the absence of continuing particle creation.

\begin{figure}
\begin{center}
\includegraphics[width=8cm,angle=0.0]{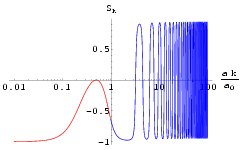}
\end{center}
\caption{Plot of the classicality parameter ${\cal S}_k$ as a function of the dimensionless variable $ka/a_{0}$ for the radiation-dominated model, for the case when the initial condition is imposed at $ka/a_0 = 1/2$. Sub- and super-Hubble regimes are color-coded blue and red respectively. ${\cal S}_k$ at late times ends up oscillating about zero with a finite [constant] amplitude. The variation of ${\cal S}_k$ reflects the complex trajectory of $z_k$, which at late times is a circle centered on the origin.}
\label{radS}
\end{figure}

This may be compared with the late time variation of the classicality parameter, which is shown in figure~\ref{radS}. ${\cal S}_k$ is oscillatory with an amplitude less than unity; this of course is a consequence of the nature of the underlying $q$-$p$ correlation, which, as can be directly figured using \eq{sig_r} and \eq{j_r}, remains finite and oscillatory over the course of the late time adiabatic evolution. Recalling the corresponding behavior in the non-adiabatic de Sitter case we dealt with earlier, the evolution, with regard to this feature, differs sharply in the two scenarios (as does the mean particle number).  

The imaginary part of the time dependent effective action at late times directly expresses the asymptotic particle content in the quantum state, and an out-vacuum is definable in the adiabatic region. On the other hand, making use of the approximate expression for $z_k$ in \eq{zr_adiab} gives the following expression for the real part of the effective Lagrangian at late times:
\beq
\textrm{Re} L_{eff} = -\frac{1}{4}\frac{m_k'}{m_k} \textrm{Im} z_k \approx  -\frac{1}{2}\frac{a_0}{a} \l[ -\frac{1}{2}\frac{a_0}{k a} + \textrm{Im} \l( {\cal R}_k e^{-2 i k \frac{a}{a_0} } - \frac{i}{2}\frac{a_0}{k a} \l( {\cal R}_k^2 e^{-4 i k \frac{a}{a_0}} + 2{\cal R}_k e^{-2 i k\frac{a}{a_0}} \r) \r) + {\cal O}\l( \frac{a_{0}^2}{k^2 a^2} \r) \r].
\eeq
Except for the first term in the brackets, the remaining terms (up to first order in $1/a$) are oscillatory in the $a \to \infty$ limit, so their contribution to the effective action may be expected to be highly suppressed. Thus, the first non-trivial contribution to the effective action from the late time adiabatic evolution comes from the non-oscillatory term with magnitude $a_{0}^2/(4ka^2)$, which is equivalent to $\epsilon^2(a) k / 16$.

Based on the analysis carried out above, combined with that of the de Sitter model, it is evident that peaking of the Wigner function on the corresponding classical phase trajectory at late times is a fairly general feature, seemingly independent of the adiabatic vs. non-adiabatic nature of the time evolution. In contrast, the classicality parameter, providing an alternative handle to quantify the variation of the Wigner distribution, shows appreciably dissimilar behavior in the two examples, and bears a rather close relation with particle production. As alluded to earlier, these observations are pretty much similar to the results obtained our in the toy models in Paper I revealed.   

The analysis of the preceding sections brings out the contrast of time evolution of quantized field modes in the inflationary scenario with that in the electric field case, and we would like to emphasize it once again here. A comparison of the plots for various quantities evaluated for the de Sitter case in section~\ref{sec:ds} with those in section~\ref{sec:ef} makes this point amply clear. In particular, a constant electric background provides adiabatic \emph{in} as well as \emph{out} regions where the notion of a particle can be unambiguously defined, which is not the case for the de Sitter model, where the late time evolution is highly non-adiabatic. Moreover, the power spectrum (related to the two-point function for the field) of field modes does not provide a direct measure of the particle content in the adiabatic limit in the electric field scenario (see \eq{ps_ef}) due to the asymptotically oscillatory nature of $z_{\bf k}$; in contrast, for the de Sitter background, as is clear from \eq{ps} such a correspondence is made possible in the late time non-adiabatic phase of evolution where $z_k \to 1$. As pointed out before, the electric field case is qualitatively much closer to a decelerating universe like the radiation-dominated example, at least in being characterized by a late time adiabatic phase. The similarities in the nature of the plots in sections \ref{sec:ef} and \ref{sec:rd} make this fairly clear.

\subsection*{4. A radiation-dominated universe going over to an asymptotically de Sitter phase}

We would like to provide another illustration of our formalism by considering quantized field evolution in a cosmological model which decelerates (radiation-dominated) initially before making a transition to a de Sitter phase of accelerated expansion. The Friedmann equation for the scale factor describing such a universe~\cite{paddy03} is given by
\beq
\l(\frac{\dot a}{a}\r)^2 = \frac{8 \pi G}{3} \l( \frac{\rho_{r 0}}{a^4} + \rho_{\Lambda} \r)
\eeq
($\rho_{r 0}$ being the radiation energy density at $a=a_0 \equiv 1$) and the solution in terms of the cosmic time $t$ has the form
\beq
a(t) = \l( \frac{\rho_{r 0}}{\rho_{\Lambda}} \r)^{1/4} \l[\sinh \l( 2 \sqrt{\frac{8 \pi G}{3}\rho_{\Lambda}}~t \r)\r]^{1/2} \equiv \l( \frac{\rho_{r 0}}{\rho_{\Lambda}} \r)^{1/4} \l(\sinh \l( 2 \tau \r)\r)^{1/2} 
\eeq
where we have introduced the dimensionless variable $\tau \equiv \sqrt{\frac{8 \pi G}{3}\rho_{\Lambda}}~t$.

The adiabaticity parameter  $\epsilon_k$ corresponding to this expansion factor diverges in \emph{both} extreme limits ($a \to 0$ and $a \to \infty$), so asymptotic \emph{in} and \emph{out} states are not definable. Nonetheless, one can solve \eq{eqz} to determine the time evolution of the quantum state of every mode. 

\begin{figure}[t]
\begin{center}
\subfigure[ ]{\label{nrl}\includegraphics[width=7.5cm,angle=0.0]{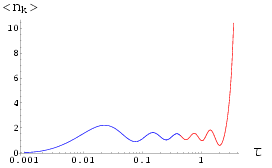}}
\hspace{0.4cm}
\subfigure[ ]{\label{zrl}\includegraphics[scale=0.7]{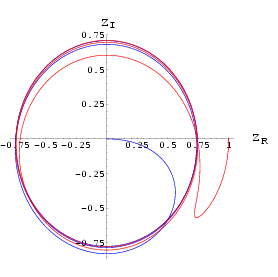}}
\subfigure[ ]{\label{Srl}\includegraphics[width=8cm,angle=0.0]{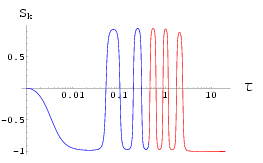}}
\end{center}
\caption{Plots of the particle number $\langle n_k \rangle$, complex trajectory of $z_k$ and classicality parameter ${\cal S}_k$ as functions of the dimensionless variable $\tau$ for the radiation+Lambda model. The instantaneous ground state is imposed at $\tau=0.001$, and radiation-dominated and Lambda-dominated phases have been color-coded blue and red respectively. $\langle n_k \rangle$ diverges at late times (super Hubble scales). $z_k$, after circling around for a while, asymptotically ends up at $z_k=1$. This is reflected in the evolution of ${\cal S}_k$, which oscillates in the intermediate sub Hubble phase before saturating at unity.}
\label{rad}
\end{figure}

Figures \ref{nrl}, \ref{zrl} and \ref{Srl} show the numerical solutions for the mean number of particles, complex path of $z_k$ and classicality parameter for a mode $k$ corresponding to $\sqrt{3/8 \pi G}k/ \l(\rho_{r 0}\rho_{\Lambda 0} \r)^{1/4} = 10 $, plotted against the variable $\tau$. Here, the initial condition $z_k = 0$ has been imposed at $\tau=0.001$ (which corresponds to $R_H/\lambda_p = 0.5$) and the radiation-dominated and Lambda-dominated phases have been color-coded blue and red respectively. The average particle number, after a short period of transient oscillations (when the mode is temporarily inside the Hubble radius) monotonically increases at late times, and this mirrors the behavior of $z_k$, which circles around the origin for a while before finally winding up at the point $z_k =1$. The behavior of ${\cal S}_k$, saturating at unity after a phase of oscillatory variation, follows from this. In fact, one can see pretty clearly that the plots for this model can be thought of as obtained by `patching up' the corresponding plots for the pure radiation-dominated and de Sitter examples which were considered earlier. Our time-dependent formalism thus provides a convenient way to visualize the differences in mode evolution during the decelerating and accelerating phases of expansion.   

\subsection*{5. Approximate analysis of accelerating models}

We would now like to verify that the features suggested by the preceding examples are fairly generic, i.e. they hold to a reasonable extent for all power law models (i.e. for all values of the index $p$). To this end, an approximate analysis of the time evolution for arbitrary $p$ is outlined below. We will first consider the range $p>1$ corresponding to inflation, followed by the case of decelerating models with $0<p<1$.

The accelerating models are described by the following scale factor~\cite{pad}:
\beq
a(t) = a_0 t^p \qquad (p>1)
\eeq
which, in terms of the conformal time coordinate can be re-expressed as
\beq
a(\eta) = (-{\cal H} \eta)^{\gamma + 1} \qquad (-\infty < \eta < 0) 
\eeq
where $\gamma = (2p-1)/(1-p)$ with $\gamma \leq -2$ and ${\cal H}=(p-1)a_{0}^{1/p}$. De Sitter inflation is a particular example belonging to this set, and corresponds to the choice $\gamma = -2$. Needless to say, all these models are characterized by non-increasing Hubble radius, and the evolution of every field mode turns strongly non-adiabatic at late times following Hubble crossing.  

We would like to consider a scenario where the quantum state of any mode starts out in the instantaneous vacuum in the asymptotic past. The solution to the equation for $\mu_k$, \eq{mueq}, which gives vanishing particle number in the $\eta \to -\infty$ limit is given by 
\beq
\mu_{k}(\eta) = N (k\eta)^{\nu}  H^{(1)}_{\nu}(k\eta)   \label{mua}
\eeq
where $N = \sqrt{\pi}/2 k^{\nu}$ and $\gamma$ has been replaced by the new variable $\nu = -\gamma-1/2$; the normalization has been chosen so that $\mu_{k}'^{*}\mu_{k} - \mu_{k}'\mu_{k}^{*} = -i/a^2$, so the Wronskian is set equal to unity. 

Fixing the form of $\mu_k$, in principle completely specifies the time evolution of the state in all respects. Since we however are particularly interested in determining the late time behavior, the $\eta \to 0$ limit of the various quantities will be considered. Let us start with the solution for $\mu_k$ in \eq{mua}. At super-Hubble scales ($k|\eta|\ll 1$), the approximate form of the Hankel function is given by
\beq
H^{(1)}_{\nu}(k\eta) ~\approx~ A \l( \frac{k\eta}{2} \r)^{-\nu} + B \l( \frac{k\eta}{2} \r)^{2-\nu}
\eeq
where
\beq
A = -\frac{i}{\Gamma(1-\nu) \sin \pi \nu} \quad,\quad B = \frac{i}{\Gamma(2-\nu) \sin \pi \nu} 
\eeq
and so
\beq
\mu_{k}(\eta) \approx N \l[ 2^\nu A + 2^{\nu-2} B (k\eta)^2 + {\cal O}(\eta^{2+\lambda})\r] \quad,\quad \mu_{k}'(\eta) \approx N \l[ 2^{\nu-1} B k^2 \eta +{\cal O}(\eta^{1+\lambda}) \r]  \qquad(\lambda > 0).
\eeq
This gives the following approximation for the energy in the mode at late times (super-Hubble scales) to the lowest order of approximation:
\beq
{\cal E}_{k} ~=~ \frac{a^2}{2} \l( |\mu_{k}'|^2 + k^2 |\mu_{k}|^2  \r)  ~\approx~ 2^{2\nu-3}\pi k |A|^2 \l( k \eta \r)^{1-2\nu} ~\equiv~ \frac{2^{2\nu-3}\pi k}{\l(\Gamma(1-\nu)\r)^2 \sin^2 \pi \nu}\l( k \eta \r)^{1-2\nu}
\eeq
Since $\nu \geq 3/2$ for this class of models (corresponding to $p>1$), the energy in every mode diverges at sufficiently late times ($k\eta \to 0$). This also means that the mean number of particles, being directly related to the energy, diverges as well:
\beq
\langle n_{k} \rangle ~=~ \frac{{\cal E}_{k}}{k} - \frac{1}{2} ~\approx~  \frac{{\cal E}_{k}}{k} ~\approx~ \frac{2^{2\nu-3}\pi}{\l(\Gamma (1-\nu)\r)^2 \sin^2 \pi \nu } \l( k \eta \r)^{1-2\nu}.
\eeq

Let us now move over to the phase space and consider the behavior of the corresponding Wigner function at late times; this is specified in terms of the functions $\sigma_k^2$ and ${\cal J}_k$, whose late time approximations have the forms
\beq
\sigma_k^2 = 2 |\mu_{k}|^2 \approx N^2 \l[ 2^{2\nu} |A|^2 + 2^{2\nu-2}(AB^* + A^*B)(k\eta)^2 \r]  \label{siga}
\eeq
and
\beq
{\cal J}_k = \l(\frac{a^2}{2}\r) \frac{1}{|\mu_{k}|^2} \frac{d |\mu_{k}|^2 }{d \eta} \approx \frac{(-{\cal H} \eta)^{2\gamma + 2}}{2} \l[ 1 + {\cal O}(k^2\eta^2) \r].   \label{ja}
\eeq
It is clear that $\sigma_k^2$ has a finite limiting value, while ${\cal J}_k$ is an increasing function of time, and diverges at super-Hubble scales. This represents a Wigner function that sharply peaks on the $p$-axis at late times.  

In contrast, in the asymptotic past as $z_k \to 0$, $\sigma_k ^2 \propto a^{-2} \to \infty$ while ${\cal J}_k \to 0$, describing a Wigner function that peaks around the $q$-axis. 

The behavior of the Wigner function needs to be compared with that of the corresponding classical trajectory in phase space. A general solution to the classical equation of motion is given by
\beq
q_k = \textrm{Re} \l[ {\cal C}\eta^{\nu}  H^{(1)}_{\nu}(k\eta) \r]   \label{ct_acc}
\eeq
where ${\cal C}$ is an arbitrary complex constant. At sufficiently early times ($\eta \to -\infty$), using the asymptotic form of the Hankel function, it can be shown that
\beq
q_k ~\approx~ \sqrt{\frac{2}{\pi k}} \textrm{Re}\l[ {\cal C} \eta^{\nu-\frac{1}{2}} e^{i(k\eta + \phi)} \r] \qquad \l(\phi = - \frac{\pi \nu}{2} - \frac{\pi}{4} \r)
\eeq
while the conjugate momentum is given by
\beq
p_k = a^2 q'_k \approx (-{\cal H})^{1-2\nu} \sqrt{\frac{2}{\pi k}} \l[ \l(\nu-\frac{1}{2}\r) \textrm{Re}\l( {\cal C} \eta^{-\nu-\frac{1}{2}} e^{i(k\eta + \phi)} \r) - k~\textrm{Im} \l({\cal C} \eta^{-\nu + \frac{1}{2}} e^{i(k\eta + \phi)} \r) \r] 
\eeq
which implies that in the asymptotic past, $p_k \to 0$ with an oscillatory $q_k$ whose amplitude sharply increases in the $\eta \to -\infty$ limit, and the classical trajectory starts close to the $q$-axis. In contrast, at late times ($k |\eta| \ll 1$), we have
\beq
q_k \approx \textrm{Re} \l[{\cal C} \l(A \l( \frac{2}{k}\r)^\nu + B \l( \frac{2}{k}\r)^{\nu-2} \eta^2 \r) \r]
\eeq
with the corresponding momentum being
\beq
p_k = a^2 q'_k \approx 2 (-{\cal H})^{1-2\nu} \textrm{Re}\l({\cal C}B \l(\frac{2}{k}\r)^{\nu-2}\r) \eta^{2-2\nu}.
\eeq
Since $\nu \geq 3/2$, $q_k/p_k \to 0$ as $\eta \to 0$, and we have a trajectory that ends up on the $p$-axis. Thus, based on the above limits one concludes that the Wigner function,for all $p>1$, peaks on the classical trajectory in \emph{both} asymptotic regions. This generalizes to all accelerating power law models what was earlier found for the de Sitter case, namely that this aspect (peaking on the classical trajectory) of the Wigner function may not provide much information regarding the approach to classicality.

Furthermore, using the approximations for $\sigma_{k}^2$ and ${\cal J}_k$ in Eqs.~(\ref{siga}) and (\ref{ja}), it trivially follows that the $q$-$p$ correlation (being just the product ${\cal J}_k\sigma_{k}^2$) will diverge as ${\cal J}_k \to \infty$ at late times, implying a classicality parameter which ends up saturating at unity, accompanying strong particle production. This feature, too, was found to occur in the de Sitter example, and now has been generalized to all accelerating models of the type $a\propto t^p$. 

We will now move on to examine the other class, that of decelerating models, and explore the generality of the features the radiation-dominated model taken up earlier suggested.     

\subsection*{6. Approximate analysis of decelerating models} 

These models are characterized by the scale factor
\beq
a(t) = a_0 t^p \qquad (0<p<1)
\eeq
and in terms of conformal time, have the form
\beq
a(\eta) = (-{\cal H} \eta)^{\gamma + 1} \qquad \l(0 < \eta < \infty \r)
\eeq
where $\gamma =  \l(2p-1\r)/\l(1-p\r)$ and ${\cal H}=(p-1)a_{0}^{1/p}$. In contrast with the situation in accelerating models, the adiabatic parameter given by $\epsilon_k = 2/k\eta$ is strongly suppressed in the late time sub-Hubble regime ($k\eta \to \infty$). 

Like in the previous case, we would like to focus on the late time behavior of various quantities associated with the evolving quantum state. So we start with the general solution to \eq{mueq}, which is given by 
\beq
\mu_{k}(\eta) = \frac{\sqrt{\pi \eta}}{2 a(\eta)} \l[ {\cal A}_k H^{(1)}_{\nu}(k\eta) + {\cal B}_k H^{(2)}_{\nu}(k\eta) \r]
\eeq
where $\nu = -\gamma -(1/2)$. It should be recalled that the $\eta \to 0$ limit here is highly non-adiabatic, so a natural in-vacuum state is not definable in this region. We will therefore just work under the assumption that the state is chosen to coincide with the instantaneous vacuum at some moment $\eta=\eta_i$, which basically fixes ${\cal B}_k/{\cal A}_k$. At sufficiently late times, however, every mode (in any arbitrary state) will eventually turn sub-Hubble and evolve adiabatically, so we have, for large values of $k\eta$,
\beq
\mu_{k}(\eta) \approx \frac{1}{\sqrt{2k}a(\eta)}\l({\cal A}_k e^{ik\eta - i\frac{\pi \nu}{2} - i\frac{\pi}{4}} + {\cal B}_k e^{-ik\eta + i\frac{\pi \nu}{2} + i\frac{\pi}{4}}\r)
\eeq
which gives the following approximation for $z_k$ at late times:
\beq
z_k(\eta) ~\approx~ {\cal R}_k e^{-2ik\eta} - i(1+{\cal R}_ke^{-2ik\eta})^2 \frac{(1-2\nu)}{4 k\eta} + {\cal O}\l((k\eta)^{-2}\r) 
\eeq
where ${\cal R}_k = \l({\cal B}_k/{\cal A}_k\r)\exp \l(i \pi \nu + i \pi/2 \r)$. Using this approximate expression, one can derive approximations for the particle number as well as the correlation parameter in the adiabatic limit:
\beq
\langle n_{k} \rangle ~\approx~ \frac{|{\cal R}_k|^2}{1-|{\cal R}_k|^2} \l[ 1 -  \frac{\l(1-2\nu\r)}{2k\eta} \frac{Im({\cal R}_k e^{-2ik\eta})}{|{\cal R}_k|^2} + {\cal O}\l((k\eta)^{-2}\r) \r]
\eeq
which implies that the particle number saturates at the finite value $|{\cal R}_k|^2/\l(1-|{\cal R}_k|^2\r) \equiv |{\cal B}_k|^2/\l(|{\cal A}_k|^2 - |{\cal B}_k|^2 \r)$, and
\beq
{\cal J}_k \sigma_k^2  ~=~ \frac{2 Im z_k}{1-|z_k|^2} ~\approx~ \frac{2 Im({\cal R}_k e^{-2ik\eta})}{1-|{\cal R}_k|^2} - \frac{\l(1-2\nu\r)}{2k\eta} \l(\frac{ Re(1+{\cal R}_k e^{-2ik\eta})^2 + 2 \l(Im({\cal R}_k e^{-2ik\eta}) \r)^2 }{1-|{\cal R}_k|^2}\r) + {\cal O}\l(\frac{1}{k^2\eta^2}\r).
\eeq
In the late time limit ($k\eta \to \infty$), only the first oscillatory term survives, and describes a correlation function that oscillates with a finite amplitude. Thus, as particle production is suppressed in the adiabatic limit, the $q$-$p$ correlation remains bounded, generalizing the behavior which was found earlier in the radiation dominated model.

From the expressions for $\sigma_k^2$ and ${\cal J}_k$ (see table), it is clear that in the strongly adiabatic limit ($k\eta \to \infty$) since $|z_k|$ does not go to unity, the behavior of these functions is primarily determined by the late time limit of $m_k \equiv a^2$; this implies $\sigma_k^2 \to 0$ while ${\cal J}_k$ is oscillatory (due to the oscillatory nature of $z_k$) with an amplitude that is proportional to $a^2$, describing a Wigner function that progressively gets peaked on the $p$-axis at late times. The Wigner function, however, was started off at $\eta=\eta_0$ in an uncorrelated state (${\cal S}_k=0$) that is fairly spread out (with $\sigma_k^2(\eta_i)=k^{-1}a^{-2}(\eta_i)$). Thus the Wigner function gets sharply peaked with time even though the mean particle number remains finite all along. As for the classical trajectory, it has the same form as in the $p>1$ models, \eq{ct_acc}; however, the late time limit here corresponds to taking $\eta \to \infty$, and this gives
\beq
q_k \approx \sqrt{\frac{2}{\pi k}} Re\l[ {\cal C} \eta^{\nu-\frac{1}{2}} e^{i(k\eta + \phi)} \r] \equiv \sqrt{\frac{2}{\pi k}} C \eta^{\nu-\frac{1}{2}} \cos\l(k\eta + \phi + c\r)
\eeq 
where ${\cal C}$ has been written as $Ce^{ic}$, and
\beq
p_k = a^2 q'_k \approx (-{\cal H})^{1-2\nu} \sqrt{\frac{2}{\pi k}} C \l[ \l(\nu-\frac{1}{2}\r) \eta^{-\nu-\frac{1}{2}} \cos \l( k\eta + \phi + c \r) - k \eta^{-\nu + \frac{1}{2}} \sin \l( k\eta + \phi + c \r) \r].
\eeq
Since $\nu < 1/2$ for $0<p<1$, we have $q_k(\eta) \to 0$ as $\eta \to \infty$, and the classical trajectory (for arbitrary ${\cal C}$) ends up on the $p$-axis, matching with the late time behavior of the Wigner distribution. As in the case of the classicality parameter above, this too is a generalization of what was seen in the radiation-dominated example. 

To conclude, based on the approximate analysis of the last couple of subsections, the two distinct late time limits of power law expansion, adiabatic vs. non-adiabatic, show clearly differing behavior as regards particle production and variation of the classicality parameter (found to be sensitive to the particle creation) providing one possible measure of phase space correlations. Peaking of the Wigner function on the classical trajectory, on the other hand, is a fairly general feature, happening whenever one of the oscillator parameters blows up/goes to zero, and is quite independent of the particle production, i.e. adiabatic vs. non-adiabatic evolution. The observations drawn here, it may be noted, are broadly similar to, and thus to an extent corroborate, what we learned from the various toy models dealt with in Paper I.  

\section{Discussion}   \label{sec:diss}

In the previous sections, we have attempted to shed some light on a few conceptual issues arising in the study of the quantized time dependent oscillator, focusing our attention on a couple of conventional examples drawn from field theory in external backgrounds that hold particular physical relevance; this has been intended as an exploitation of the analysis (based on the Schrodinger picture formalism) that was carried out in Paper I. A quick recapitulation of the key features of the time evolution (of a given fourier mode) in the asymptotic limits, as revealed by our analysis of the examples of de Sitter and radiation models and the constant electric background case, is tabulated below:

\begin{center}
  {\scriptsize
\begin{tabular}{|c|c|c|c|}
  \hline
   & & & \\
 REGIME OF & ~~de Sitter model~~  & ~~Radiation dominated model~~  & ~~Constant Electric field~~
  \\
INTEREST& $a(\eta)=-1/H\eta$ & $a(\eta)=a_{0}\eta$ & $A_{z}(t)=-Et$ \\
&  &  &  \\
\hline

& & & \\

& $\eta \to -\infty$ & ~$\eta \to 0$ (But initial condition is ~ & $t \to -\infty$ \\
 

& & chosen here at $\eta = \eta_{0} \neq 0$) & \\

& & &\\

  & Adiabatic & Highly non-adiabatic & Adiabatic \\

& & & \\

 ~~EARLY TIMES~~ & $\langle n_k \rangle \to 0$ & $\langle n_k \rangle = 0$ & $\langle n_{\bf k} \rangle \to 0$ \\

& & & \\

& $\sigma_k^{2} \to \infty$ & $\sigma_k^{2}$ : finite  & $\sigma_{\bf k}^{2} \to 0$ \\

& & & \\

 & ${\cal J}_k \to 0$ & ${\cal J}_k = 0 $ & ${\cal J}_{\bf k} \to 0$\\

& & & \\

   & ${\cal S}_k \to 0$ & ${\cal S}_k = 0$ & ${\cal S}_{\bf k} \to 0$\\

& & & \\

\hline

& & & \\

& ~~~Strongly non-adiabatic~~~ & Adiabatic & Adiabatic \\

& & & \\

& $\eta \to 0$ & $\eta \to \infty$ & $t \to \infty$ \\

& & & \\

 & $\langle n_k \rangle \to \infty$ & $\langle n_k \rangle \to $ constant & $\langle n_{\bf k} \rangle \to $ constant\\

& & & \\

~~LATE TIMES~~ & $\sigma_k^{2} \to$ finite & $\sigma_k^{2} \to 0$  & $\sigma_{\bf k}^{2} \to 0$ \\

& constant & & \\

 & & & \\

 & ${\cal J}_k \to \infty$ & ${\cal J}_k$ : oscillatory with & ${\cal J}_{\bf k}$ : oscillatory with \\

& & increasing amplitude & increasing amplitude \\

& & & \\

   & ${\cal S}_k \to 1$ & ${\cal S}_k $ : oscillates about ${\cal S}_k=0$ & ~~${\cal S}_{\bf k} $ : oscillates about ${\cal S}_{\bf k}=0$~~\\

& & with fixed amplitude & with fixed amplitude\\

& & &\\

\hline

\end{tabular}
}
\end{center}
\vspace{0.3cm}   

For the examples (particularly those of the de Sitter and electric field models) considered here, the Wigner function for any field mode is found to get peaked on the corresponding classical phase space trajectory not just at late times, but even in the asymptotic past. This renders the interpretation of classicality based solely on peaking on the classical trajectory dubious, and naturally brings in the need to have an additional measure to specify the degree of correlation, if one seeks to address the question of approach to classicality confining oneself solely to the phase space evolution of the system. Our choice of the classicality parameter, which is found to closely track the particle production, proves fairly reasonable in this regard and is seen as having a more sensible interpretation in some situations, like being zero in a vacuum state even though the Wigner function may peak in this limit etc..  

Let us reconsider the various limiting cases we came across in our examples in a step-by-step manner. First, let us consider the limit of early times. For all three cases considered, the function ${\cal J}_{\bf k}$ as well as the classicality parameter ${\cal S}_{\bf k}$ vanish at the moment when the mode is started off in the instantaneous ground state, and this clearly represents an uncorrelated Wigner distribution, although it is peaked in a limited region in phase space. (In the de Sitter and electric field cases, one can start off in the instantaneous vacuum state at $t=-\infty$, but for the radiation dominated universe, a non-zero initial time $\eta_i$ is chosen.)

As for the late-time behavior, the three examples differ in specific respects. For the de Sitter model, $\sigma_{\bf k} ^{2}$ approaches a finite limiting value while ${\cal J}_{\bf k} $ blows up. It can also be seen that the classicality parameter smoothly goes to unity in the $a \to \infty$ limit [while it remains nearly zero at sub-Hubble scales in the asymptotic past]. Clearly, these features are indicative of a fairly correlated phase space distribution, and this is in line with the usual conclusion of classicality emerging on super-Hubble scales \cite{infl2,pad,wig,class}.
For the radiation-dominated case, the Wigner distribution ends up getting peaked on the $p$-axis (coinciding with the corresponding classical trajectory) at late times on account of the fact that $\sigma_{\bf k} ^{2} \to 0$. 
 
As regards the variation of the classicality parameter, it turns out that this measure is zero to start with (at the instant when the mode is in the instantaneous ground state) but subsequently (for $\eta > \eta_0$) develops an oscillatory profile that has a finite amplitude at late times. So, it can probably be argued that in some sense, the state gets significantly more correlated {\it in comparison with} the initial state, although this is in contrast with the de-Sitter case [where ${\cal S}_{\bf k} \to 1$ at late times], and is not so obvious. The difference is clearly tied with the adiabatic vs. non-adiabatic nature of the late time evolution.    

Finally, for the electric field example, the late time variation qualitatively proceeds along lines similar to the radiation-dominated model, in that $\sigma_{\bf k} ^{2} \to 0$, representing a Wigner function peaked on the $p$-axis [and which coincides with the classical trajectory]. However, the classicality parameter exhibits a variation that's more straightforward to interpret: it vanishes in the asymptotic past (${\cal S}_{\bf k} \to 0$ as $t \to -\infty$) but is oscillatory with a finite, nearly constant amplitude in the $t \to \infty$ limit. So in terms of ${\cal S}_{\bf k}$, the phase space correlations clearly appear to grow more significant in the far future in relation to early times. The de Sitter and electric field cases thus display a fairly unambiguous emergence of phase space correlations with time, but the evolution in the radiation model, in this particular respect, is a little less clear.   
The bottom line, thus, is that particle production is not necessary for peaking of the Wigner function on the classical trajectory as the three examples that were considered indicate, where this was found to happen $\it even$ as $\langle n_{\bf k} \rangle$ saturated at late times or decreased to zero in the asymptotic past. (It may however be noted that one or the other of the oscillator parameters blows up/goes to zero in these limits.) Our analysis thus points toward the fact that peaking on the classical trajectory \emph{alone} need not suffice to regard a system as turning classical; one needs to supplement this criterion with looking at some sensible measure of phase space correlations to settle this question. The classicality parameter that we have used is a simple yet useful construct in this regard, and leads one to conclude that strong particle creation, although not necessary for causing the Wigner function to get concentrated on the classical trajectory, is a necessary condition for driving the quantum state to become highly correlated (i.e. $\langle q p \rangle \to \infty$).

\section*{ACKNOWLEDGEMENTS}

G.M. is supported by the Council of Scientific $\&$ Industrial Research, India.

\end{document}